%% file: main.tex
\documentclass[a4paper, 12pt]{article}
\usepackage[utf8]{inputenc}
\usepackage{amsmath,amsthm,amssymb,times,geometry,color}
\usepackage{natbib}
\usepackage{graphicx}
\usepackage{listings}
\usepackage{enumerate}
\usepackage{array}
\usepackage{adjustbox}
\usepackage{commath}
\usepackage{multirow}
\usepackage[colorlinks=true,linkcolor=blue,citecolor=blue]{hyperref}
\usepackage{array}
\usepackage{float}
\usepackage{comment}
\usepackage{setspace}
\usepackage{mathptmx}
\usepackage{subfig}
\usepackage{bm}
\usepackage{lipsum,pdflscape}
\usepackage{threeparttable}
\usepackage{makecell}
\usepackage{booktabs}
\usepackage{tabularx}
\usepackage{placeins}
\usepackage{authblk}
\usepackage[title]{appendix}
\begin{document}

\renewcommand\Authands{, } 
\thispagestyle{empty}
\title{Long-term Health and Human Capital Effects of Early-Life Economic Conditions \thanks{We gratefully acknowledge financial support from the European Research Council (ERC) under the European Union's Horizon 2020 research and innovation programme (grant agreement no. 851725). This research has been conducted using the UK Biobank Resource under Application Number 74002. We thank Paul Hufe and Jeremy McCauley for comments and suggestions on an earlier version of this paper. We also thank seminar participants at the European Social Science Genetics Network Conference, the Essen Health Conference, the Scottish Economic Society Conference and the University of Bristol for their helpful comments and suggestions.}}

\author[1,2]{Ruijun Hou}
\author[2]{Samuel Baker}
\author[2,3,4]{Stephanie von Hinke}

\author[5,2,4]{Hans~H.~Sievertsen}
\author[2]{Emil Sørensen}
\author[2]{Nicolai Vitt}

\affil[1]{\small {Department of Population Health Sciences, University of Leicester, Leicester, United Kingdom.}}
\affil[2]{School of Economics, University of Bristol, Bristol, United Kingdom}
\affil[3]{Institute for Fiscal Studies, London, United Kingdom}
\affil[4]{Institute for the Study of Labor (IZA), Bonn, Germany}
\affil[5]{The Danish Center for Social Science Research,VIVE}

\date{\today }

\maketitle
\begin{abstract}

\singlespacing
\noindent 
We study the long-term health and human capital impacts of local economic conditions experienced during the first 1,000 days of life. We combine historical data on monthly unemployment rates in urban England and Wales from 1952 to 1967 with data from the UK Biobank on later-life outcomes. Leveraging variation in unemployment driven by national industry-specific shocks weighted by industry's importance in each area, we find no evidence that small, common fluctuations in local economic conditions during the early life period affect health or human capital in older age.

\vspace{3mm}

\vspace{1mm}\newline
\end{abstract}

\doublespacing

\maketitle 

\newpage
\pagenumbering{arabic}
\clearpage

\section{Introduction}
The first 1,000 days of a child's life -- from conception till their second birthday -- are considered some of the most critical for child development \citep[see e.g.,][]{black2013maternal, HSCC20119}. A large literature shows that exposure to adverse conditions during this phase can have life-long consequences \citep[for reviews, see e.g.,][]{almond2011killing, Almond2018}. The bulk of this evidence is based on large negative shocks such as famines or pandemics \citep[e.g.,][]{RN846, RosalesRueda2018, beach2022} or targeted policies such as medical interventions or health/educational programs \citep[e.g.,][]{belfield2006high, almond2010estimating, wust2022universal}. We know considerably less however, about the long-term consequences of small and common variations in conditions at birth, especially on human capital outcomes in adulthood \citep{currie2011human}. 

In this paper, we study the long-term impacts of local economic conditions in early life on individuals' outcomes in late adulthood. We use the unemployment rate in one's local area of birth during the first 1,000 days of life to proxy for such conditions. 
On the one hand, one would expect an increase unemployment to reflect a deterioration in the early-life environment. For example, they can impact household resources (e.g., by affecting parental employment), which in turn can affect child development \citep{del2012intrafamily,bono2016early}. Similarly, they may affect parental stress or child nutrition, with potential longer-term impacts on child development \citep{persson2018family, von2022mental}. Adverse local economic conditions may also impact public resources that in turn affect individuals' development through their effects on local health or educational programs \citep{Keegan2013, Jackson2021}.
On the other hand, however, there is a large literature suggesting that adverse economic conditions may improve one's health, driven by increased health investments \citep[e.g., more doctor visits, reduced risky health behaviours, healthier diets; see e.g.,][]{Ruhm2000, ruhm2003good}. This, in turn, could have positive inter-generational spillovers, improving the health and well-being of offspring \citep{von2014alcohol, nilsson2017alcohol, pereira2022interplay}. 

Whether fluctuations in local economic conditions in the early childhood period have long-run effects on individuals in older age is therefore an empirical question. Estimating the extent to which variation in later-life outcomes is driven by early-life circumstances is important not only to improve our understanding of human capital development, but also because it allows us to gauge whether the total impact of adverse economic conditions differs from any contemporaneous effects. 

We use monthly variation in local, post-war labour market conditions across urban England and Wales during a period of general growth and low unemployment, allowing us to explore the impact of small and common variations in economic conditions on later life outcomes. We merge these data with observational data from the UK Biobank: a population-based cohort study of $\sim$500,000 individuals. We investigate the long-run health and human capital effects of local economic conditions during the first 1,000 days of life, where we distinguish between the prenatal period and the first and second year of life (i.e., pregnancy, age 0 and age 1). Our variable of interest may be correlated with other factors that also affect child development. Changes in child investments and demand for related products (e.g. baby formula) and services (e.g. childcare, healthcare) may drive fluctuations in local economic conditions. Similarly, public policies (e.g. expansion of early life healthcare) may affect both child development and local employment, making causal inference challenging.

To address this, we isolate variation in unemployment driven by national industry-specific unemployment rates and local (predetermined) industry shares. This essentially assumes that, for example, a national unemployment shock to the manufacturing industry will primarily impact areas with a high (predetermined) share of employment in manufacturing; it will have little impact on areas where few people work in this industry. The idea behind our IV specification is similar to a shift-share-type approach (though with a number of important differences), with the identification coming from the national industry-specific shocks; we return to the identifying assumptions in more detail below. The advantage of our IV approach is that we isolate variation in economic conditions that is unrelated to local demand for products and services, as well as local policies, which might affect both unemployment and child development. 

The contributions of our paper can be summarized as follows. First, we estimate the impacts of local economic conditions on health as well as human capital outcomes, whilst -- as highlighted by \citet{almond2011killing} -- most of the existing literature has only focused on the former. Furthermore, much of the existing literature estimates the effects of large shocks early in life, such as famines or major recessions \citep[see e.g.,][]{Thomasson2014, Stuart2022}, with relatively little evidence on the long-term impacts of frequent but much smaller shocks. While evidence on large shocks is important, the events are infrequent and less relevant for policy compared to small and common fluctuations in economic conditions that occur more regularly. 

Second, we undertake our analyses in the historical context of England and Wales in the 1950s and 1960s, which is characterised by a general lack of local-level within-year data.\footnote{Existing studies using historical data have focused on \textit{national} or \textit{county}-level data that is usually reported annually or -- in some cases -- quarterly \citep[see e.g.,][exploring medical and health data]{SMALLMANRAYNOR2003396, Mina2015, smallman2015, YuchangWuGWAS2022}, concealing much of the potentially interesting temporal and spatial variation. Local economic data for this time period are almost non-existent.} We overcome this by using newly digitised monthly historical records on local, urban and national industry-specific employment and unemployment counts between 1952 and 1967. 
In contrast to the existing literature that shows lasting consequences of large early-life economic shocks, our findings suggest that small fluctuations in local labour market conditions do not affect individuals' outcomes in older age. Our results suggest that, during periods of general economic growth and relative prosperity, such early-life exposures do not shape people's health and development.

Due to the limited spatial coverage of the monthly unemployment data, our estimation sample only covers individuals born in \textit{urban} areas during those years. However, we explore the generalizability of our findings to all areas of England and Wales. Indeed, although we only observe our variable of interest -- the local unemployment rates -- for urban areas, we observe the instrument -- the weighted industry unemployment -- for both urban and rural areas. We are therefore able to compare the reduced form results for our (smaller and urban) estimation sample with those for UK Biobank participants born in all areas of England and Wales, finding similar results across the two samples. 

Third, the availability of monthly unemployment rates additionally allows us to study potential heterogeneity in the \textit{timing} of exposure to local economic conditions, contrasting exposures that are specific to the prenatal, infancy, and very early childhood period (i.e. pregnancy, age 0 and age 1). We find little evidence of heterogeneity, with most estimates being sufficiently close to zero to rule out any strong effects.

Fourth, with our main results suggesting little impact of local economic conditions on later life health and human capital, we investigate whether these null effects conceal any heterogeneity in the impacts across individuals' characteristics. More specifically, we examine heterogeneity along three important dimensions. 

We start by investigating whether the impact of local economic conditions differs across individuals with different genetic ``predispositions'' for the outcome. We are interested in genetic heterogeneity for two reasons. One, understanding whether common fluctuations in childhood economic conditions differentially affect population subgroups is informative about the potential impacts on (arguably unfair) genetic inequalities in the population. Furthermore, since genetic variation is passed on from one generation to the next, any such effects may be reinforced through intergenerational transmission. Two, genetic heterogeneity in the effect of interest speaks to the literature on health and human capital production. This literature discusses the importance of health and human capital ``endowments'', but also emphasizes the difficulty in its measurement. Genetic variation has recently been proposed as a novel way to proxy for such endowments \citep{Muslimova2020}, given that they are fixed at conception and randomly allocated within families.\footnote{For example, it can be used to test the existence of complementarities between endowments and investments in the production of human capital \citep[][]{cunha2007technology}. Such analyses would ideally use parent-child trios or sibling data to exploit exogenous variation in genetic endowments \citep[see e.g.,][]{Muslimova2020, berg2023early}; see \citet{biroli2025economics} for a detailed discussion on using genetic data to estimate gene-environment interplay. We do not interpret our estimates as capturing such complementarities, since our measure of endowments is not exogenous. } 
We use molecular genetic data to construct so-called polygenic indices (PGIs, also known as polygenic scores or PGSs) for each outcome of interest, which can be interpreted as the best linear genetic predictor \citep{Mills2020}. Our results provide little support for genetic heterogeneity in the long-term effects of early childhood economic conditions.

We next explore whether the null effect conceals differences in effects for men and women. We find no such evidence. Finally, we investigate heterogeneity by social class. In the absence of data on (parental) socio-economic status at the time of childbirth, we characterize the socio-economic composition of the local area of birth using data on social class composition from the (predetermined) 1951 Census. Our findings provide little evidence of heterogeneous effects. 

In summary, our analysis shows no long-term effects of small, common fluctuations in local economic conditions during the first 1,000 days of life on individuals' health and human capital in later life. These could reflect true null results, but another possibility is that the null findings are driven by selection bias, selective fertility, or selective mortality. For example, if adverse economic conditions reduced years of education and simultaneously increased mortality among those affected, we may not detect the educational impact in our analysis. Similarly, if adverse conditions affect child growth but also reduce fertility, this may conceal any effects on adult height. Indeed, it may be that the households having children during times of worse local economic conditions were systematically different from households having children in better times.  Finally, if adverse economic conditions cause long-term health impairments and those of ill health are less likely to participate in the UK Biobank, then this may harm our ability to detect any long-term effects.

We explore the importance of these selection issues in more detail by estimating the impact of economic conditions on (i) selection into the UK Biobank and (ii) on local historical birth and infant death rates. Our analyses suggest that neither selection into the data, nor fertility and mortality selection, play a role in our setting and thus are not driving our null results. 

The rest of this paper is structured as follows. Section~\ref{sec:method} discusses the empirical specification, with Section~\ref{sec:data} describing the data. Section~\ref{sec:results} presents the results, followed by an analysis of potential selection bias in Section~\ref{sec:selection}. Section~\ref{sec:conclusion} concludes.

\section{Empirical specification}\label{sec:method}
To identify the effect of exposure to economic conditions during the first 1,000 days of a child's life on their later life health and human capital, our starting point is the following specification:

\begin{equation}
    Y_{itz}= \beta_a UE_{itz}^{age=a} + \bm{\gamma}' \bm{X}_{i} + \phi_t + \delta_{z}+ \mu_{z} \times t + \varepsilon_{itz} \qquad a=-1, 0, 1
    \label{equ: FE2}
\end{equation}
where $Y_{itz}$ is the outcome of interest (e.g., years of education, log wages) for individual $i$, born at time $t$ and local area $z$, and $UE_{itz}^{age=a}$ represents the average unemployment rate that the individual faced at age $a$, where $a$ refers to the prenatal (intrauterine) period ($a=-1$), the first year of life ($a=0$), or the second year of life ($a=1$, i.e., until the child is 24 months old).\footnote{We do not observe the gestational age at birth, only the year and month of the individuals' birth. We therefore assume that all pregnancies last nine months. The prenatal $UE_{itz}$ is therefore defined as the average unemployment rate during the nine months prior to birth.} 
Our parameter of interest is $\beta_a$, capturing the impact of local economic conditions at age $a$ in an individual's area of birth on their later life human capital and health outcomes.  

Our specification controls for a vector of covariates $\bm{X}_{i}$, including gender and month-of-birth dummies, where the latter capture seasonality in the outcomes of interest. Additionally, we include year-of-birth fixed effects $\bm{\phi}_t$, area-of-birth fixed effects $\bm{\delta}_{z}$ and area-of-birth-specific linear trends $\bm{\mu}_z \times t$, each of which may be correlated with both local unemployment rates and child development. The year-of-birth fixed effects capture cohort differences in the outcomes and account for general macroeconomic shocks; the area-of-birth fixed effects control for time-invariant, cross-sectional variation in unobserved characteristics such as the socio-economic composition of the area; the area-of-birth specific trends capture differential development of (e.g.) wealthier and poorer areas over time. This is important, since it is well-known that some industries were in systematic decline around this time (e.g., coal-mining), so unemployment rates in areas with a larger share of such industries are likely to be trending differently compared to the unemployment rates in other areas. $\varepsilon_{itz}$ is the error term. We cluster the standard errors by area of birth throughout. 

Even after controlling for our set of covariates and fixed effects, the estimate of $\beta_a$ can be biased for two reasons. First, variation in local economic conditions might be driven by local policies that also affect conditions for children. For example, a local policy aiming to improve healthcare or childcare quality leading to the construction of new daycare institutions and hospitals, would likely both affect child development and the local labour market.  In that case, $\beta_a$ would capture the effect of improved local economic conditions as well as the effects of improved hospital or childcare quality. Second, there is likely to be measurement error in local economic conditions, leading to attenuation bias in conventional OLS estimates.

To address these endogeneity concerns, we instrument local economic conditions using national industry-specific unemployment weighted by predetermined local industry employment shares. This approach is based on the idea that local labour markets are differentially exposed to exogenous changes in labour demand across industries depending on the local employment share in these industries. Hence, our instrument exploits plausibly exogenous variation in national industry-specific unemployment rates, interacted with area-specific industry employment shares prior to our sample period. The approach exploits the fact that areas that specialize in a given industry are more vulnerable to changes in national unemployment in that industry compared to areas that do not specialize in that industry. For example, any shock to the mining industry is expected to have affected the labour markets in mining areas such as Lancashire and Yorkshire more than those in Portsmouth or Plymouth. Our instrument therefore exploits both time and spatial variation in unemployment. Changes in national industry-specific unemployment capture the \textit{time} variation in unemployment shocks, and industry shares capture \textit{spatial} variation in the exposure to these shocks. Importantly, the variation in unemployment extracted from this instrument should be unrelated to the variation in unemployment coming from local initiatives that might be correlated with conditions affecting child development. 

We estimate an instrumental variables model, denoting the instrument for local unemployment rates as $Z_{itz}^{age=a} = \sum_k s_{kz,1951} \times UE_{itk}^{age=a}$, where $s_{kz,1951}$ denotes the share of employment in industry $k$ in area $z$ prior to our sample period (i.e., in 1951), and $UE_{itk}^{age=a}$ represents the national unemployment rate for industry $k$ at age $a$ of child $i$ born at time $t$. The first-stage equation can be written as:
\begin{equation}
    UE_{itz}^{age=a}=\alpha_a Z_{itz}^{age=a}+\theta X_{i}+\psi_t+\eta_z+\kappa_z \times t + v_{itz}
    \label{equ: first-stage}
\end{equation}

where parameter $\alpha_a$ captures the impact of the weighted shocks at age $a$ ($Z_{itz}^{age=a}$) on the observed area-level unemployment rate at the same age. 

This instrumental variable approach is related to shift-share-style designs, with the literature emphasizing the importance of discussing the source of identifying exogenous variation.\footnote{For an initial discussion of shift-share instruments, see \citet{RN887}. In addition to those cited in the text, more recent discussions of identification and inference include \citet{adao2019shift}, and \citet{de2023more}.} \citet{borusyak2022quasi} and \citet{goldsmith2020bartik} show that exogeneity of either the common shocks or the local exposure shares is sufficient for a valid identification. Indeed, using a shift-share instrument is numerically equivalent to treating either the shocks or the shares as instruments. 

In our setting, the time-invariant initial industry shares are likely correlated with local wage levels and area characteristics. While this can be addressed by controlling for area-fixed effects, the industry shares may still be endogenous if they are related to future changes in area-level unobservables. For example, industry shares may be correlated with future changes in the education and healthcare funding an area receives from the national government.

Instead of relying on exogenous industry shares, we argue that the national industry-specific levels of unemployment are plausibly exogenous to individual outcomes. Exogeneity of industry-specific national unemployment rates requires that they were not driven by local circumstances but rather by national or international economic factors. As we show below, employment in the industry groups that we consider is generally spread across a large number of areas. We argue it is therefore reasonable to assume that the shocks were not caused by local factors, but instead were driven by national or international shocks to different industries. 

Relative to the canonical shift-share design with exogenous shocks that is presented in \citet{borusyak2022quasi}, our setting differs in several ways. First, we do not observe exogenous changes in labour demand. Our observed changes in national unemployment reflect both labour demand and labour supply changes. However, as discussed in \citet{borusyak2022quasi}, aggregate unemployment can be seen as a noisy estimate of changes in labour demand and the resulting bias will be small in a setting such as ours where industries are spread across a large number of areas.\footnote{A leave-one-out adjustment which is often used to address these concerns is not feasible with our data. However, such an adjustment is not crucial in our setting as the influence of a single area on the national industry-specific unemployment is likely to be negligible.} 

Second, we do not observe all industries and therefore the original exposure shares do not sum to one, but we reweight these to ensure they add up to one. Although variation in the sum across areas may reflect differences in the importance of unobserved industries, causing the instrument to potentially leverage non-experimental variation \citep{borusyak2022quasi}, this does not pose a threat to our identification. Indeed, since our industry shares (and their sum) are time-invariant, the area fixed effects control for differences in shares of unobserved industries across areas.

Third, we use panel data of unemployment levels rather than a cross-section (of changes). In our estimations, we include area fixed effects to capture time-invariant differences between areas, and year fixed effects to capture period-specific unobservables. The inclusion of these fixed effects is furthermore aimed at isolating within-area and within-period variation in the instrument.

So far we have discussed only one of the instrumental variables assumptions, exogeneity. In addition to exogeneity, we require the instrument to affect local unemployment rates (relevance); the instrument to only affect our outcome through local unemployment (exclusion), and the instrument to have a monotonic effect on local unemployment (monotonicity). 

We test the relevance assumption empirically below. The exclusion restriction requires the weighted unemployment level to only affect child development through local unemployment. This would be violated if, for example, policymakers react to a general downturn in some industries by increasing investments in child- and healthcare facilities in areas more affected by these shocks. However, such policy responses should be captured by the area-specific time trends. 

Finally, with our instrumental variables approach we identify the local average treatment effect of changes in local unemployment rates driven by these national industry-specific shocks to unemployment. This interpretation further requires the monotonicity assumption to be satisfied. A predicted higher \textit{national} unemployment rate should always lead to higher \textit{local} unemployment rates. If policies are implemented to compensate for certain national shocks, and these policies more than compensate for the negative shock, it could lead to lower unemployment rates and a violation of the monotonicity assumption. In our setting this would, however, require such a mechanism to go beyond the area fixed effects, the year fixed effects and the local area-specific trends.

\section{Data}\label{sec:data}
We combine three data sources in our main empirical analysis: unemployment data from the \textit{Labour Gazette}, industry employment shares from the 1951 Census, and individual-level data from the UK Biobank. We here describe the data sources and how we construct the variables that we use in the main analysis. 

\subsection{Local unemployment rates}
Our main exposure of interest is one's early-life economic conditions, which we proxy by the local unemployment rate, obtained from the \textit{Labour Gazette} \citep{LabourGazette}; a historical, monthly publication by the UK Ministry of Labour. We obtain unemployment figures for 310 urban ``Travel-to-Work-Areas'' (hereafter ``areas'') across England and Wales from \citet{BakerHinke2025}. We construct each area's monthly unemployment rate as the unemployment count divided by the area's labour force, where the latter is defined as the sum of the employment and unemployment counts. To study the impact of early-life economic conditions, we focus on the first 1,000 days of life and allow for the possibility that different periods -- in utero (0-9 months before birth), infancy (0-12 months after birth), and early childhood (12-24 months after birth) -- may have differential effects. We thus calculate the average local unemployment rate for each of these periods for every possible year-month and area of birth. We then merge these averages with our individual-level outcome data (described below) using individuals' year-month and area of birth.

\subsection{The instrument}
To construct our instrument, we require national monthly industry-specific unemployment rates (i.e., the ``shifts'' in our shift-share-type approach) as well as area-level industry shares (the ``shares'') prior to our sample period. 
We digitise monthly national industry-specific employment and unemployment counts for Great Britain from the \textit{Labour Gazette}, covering the years 1951--1968.\footnote{For some industries, we only observe \textit{annual} employment counts; we use linear imputation to obtain their \textit{monthly} counts. These are Agriculture, Fishing, Professional, Transport, Finance, and Miscellaneous services.} We define the ``shift'' -- the national industry-specific unemployment rate -- as the ratio of the industry-specific unemployment counts to the industry-specific labour force, the latter defined as the sum of the employment and unemployment counts. 
To calculate the ``shares'' -- i.e., the area-specific proportion of individuals employed in each industry -- we divide the area-level industry-specific employment counts by the area-level total employed population, both obtained from the 1951 UK Census \citep[Vision of Britain;][]{VisionOfBritain}. 

Our instrument is then defined as the product of the national industry-specific unemployment rates and the area-level predetermined industry shares summed over all 18 industries observed in both  ``shift'' and  ``shares''. Note that the 1951 Census and the \textit{Labour Gazette} use slightly different industry names; the matched data therefore excludes nine industries.\footnote{\autoref{tab:Matched industry names} in \autoref{sec:AppA} details how we match the industry names from the Census data to the industry names from the \textit{Labour Gazette}.} This causes the area-specific industry shares not to sum to one due to some industries only being reported in either the \textit{Labour Gazette} or in the 1951 Census. We deal with this by re-weighting the shares to ensure they add up to one in our analysis.

Similar to above, we then calculate the average ``weighted'' unemployment rate over distinct intervals for each location and year-month. More specifically, we separately construct the instrument averaging over periods of nine months (i.e., to link to the pregnancy period), and 12 months (i.e., to link to individuals' first and second year of life), and we merge these into the individual-level UK Biobank data using individuals' location, and year-month of birth.

\subsection{Later-life health and human capital outcomes}
The UK Biobank is a prospective population-based cohort that consists of approximately half a million participants \citep{sudlow2015uk}. At the time of recruitment in 2006-2010, individuals were 40-69 years old and lived in England, Scotland, or Wales. An advantage of the data is the availability of rich information on socio-demographic characteristics, lifestyle and health-related factors. Based on the existing literature that highlights the long-term impacts of early life conditions, we select two sets of outcomes of interest: human capital and health. For the former, we focus on total years of education, hourly wages, and fluid intelligence, which have been shown to be critical components of individuals’ socio-economic development \citep{Harmon2003, Conger2010, Kyllonen2017}. For the latter, we explore the effects on height; a well-established indicator for one's standard of living, positively correlated with health and longevity \citep{Thomas2002, Samaras2003, Behrman2004, Psacharopoulos2004, Perkins2016}. 

We define individuals' years of education using the highest qualification achieved \citep[using the mapping described in][]{berg2023early}. Hourly wages are imputed based on individuals’ occupations \citep{Kweon2020}; we take their natural logarithm in the analysis. Fluid intelligence assesses the capacity to solve verbal and numerical problems and is measured by thirteen independent questions; we standardise the full score to have mean zero and standard deviation one in the analysis. Standing height (in centimetres) was measured during the interview following a standardised protocol. 

A drawback of the UK Biobank is that very limited information on individuals' early life conditions has been collected. In particular, we have no direct measures of economic conditions that individuals were exposed to. To deal with this, we use participants' area and year-month of birth to link individuals to the newly constructed unemployment data.

\subsection{Sample selection}
Prior to merging the individual-level data from the UK Biobank with the area-level unemployment data, we select our estimation sample as follows. The largest sample selection occurs when we restrict the sample to participants born between 1952 and 1967, dropping 280,534 participants. Starting our analysis sample in 1952 allows us to treat the industry shares from the 1951 Census as predetermined. Ending the sample in 1967 is because the national employment counts are only observed until 1968, when our last cohort is age one. Since the UK Biobank includes only a few participants born after this year, this is not a strong restriction. We next exclude participants born outside the UK or with missing information on their location of birth, dropping an additional 29,626 individuals. A further sample selection occurs when we drop individuals who are born in areas that are not covered by the area-level unemployment data (i.e., rural areas), excluding 64,088 participants. Finally, 26 individuals are excluded because we do not observe any of the outcomes of interest. Depending on the outcome, our final sample comprises 54,614--110,740 individuals, covering 218 of the 310 areas we have local unemployment data for.

\subsection{Descriptive statistics}
\autoref{tab: summary} reports the summary statistics for our set of outcomes, where column (1) shows the descriptive statistics for the complete UK Biobank sample, and column (2) presents the statistics for our estimation sample. The table shows that 45\% of the individuals in our estimation sample are male. On average, individuals have completed just over 13 years of education, earn an hourly wage of almost £15, have a (raw) fluid intelligence score of 6, and a standing height of 170 cm. While the differences between the full UK Biobank sample and our estimation sample are small, they suggest that the estimation sample is slightly positively selected (i.e., more education and higher wages, fluid intelligence, height). One explanation for this is that the estimation sample only includes participants born between 1952 and 1968. The individuals in the estimation sample are therefore younger than in the original sample.

The average monthly unemployment rate is around 1.6\%; the average for the instrument -- the national industry-specific monthly unemployment rate weighted by the area's industry shares -- is slightly lower at 1.4\%. 
\input{tables/summary_stat}

\begin{figure}[ht!]
 \begin{center}
   \includegraphics
    [width=0.8\linewidth]
    {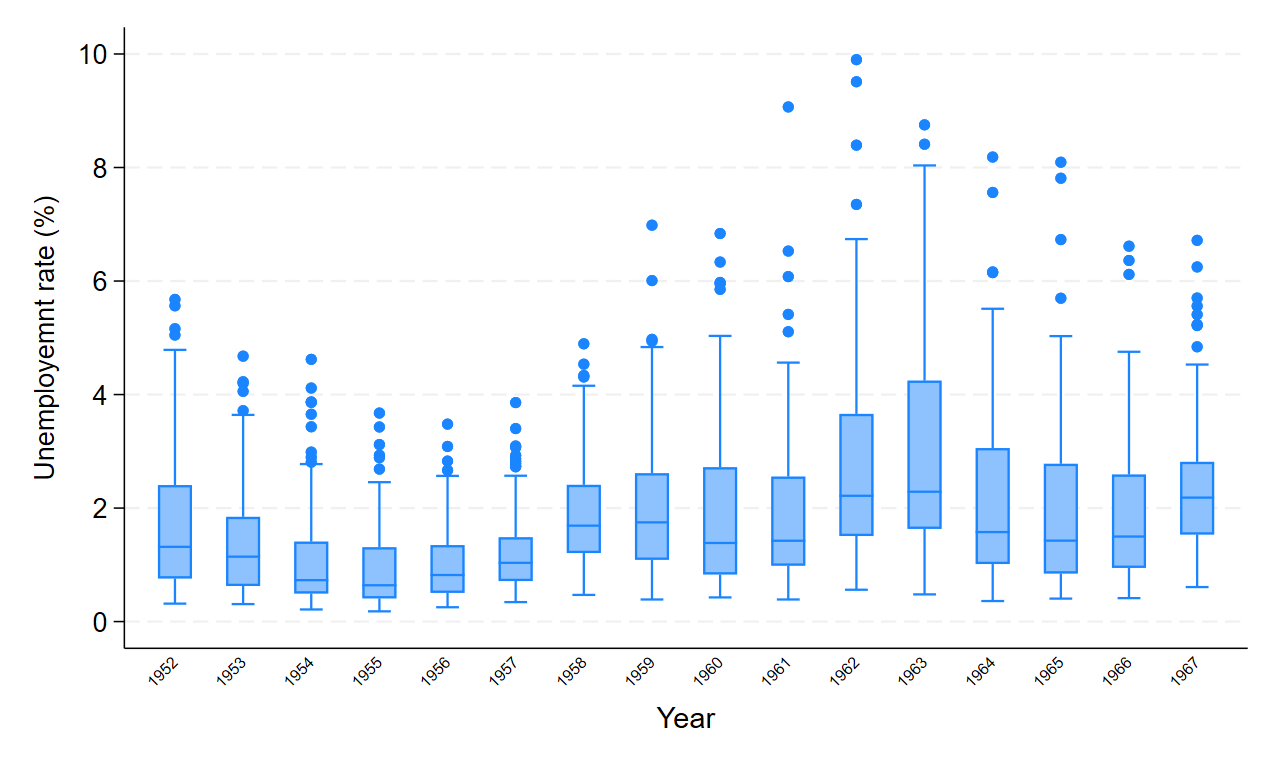}
    \caption{Annual variation in area-level unemployment}
    \begin{minipage}{0.6\linewidth}
    \hfill \vspace{6pt}\\
    \footnotesize{Notes: The boxes show the inter-quartile range of unemployment rates across 218 areas; the horizontal line inside the box shows the median, the lower line in the box is the 25th percentile and the upper line is the 75th percentile. The whiskers extend from the box by 1.5 times the inter-quartile range or to the minimum/maximum value (if this is nearer), the dots show any values beyond the whiskers.}
    \end{minipage}
    \label{fig:yearly zone ue}
 \end{center}
\end{figure}

To examine the variation in unemployment rates during our observation period, \autoref{fig:yearly zone ue} presents a box plot of area-level annual unemployment rates between 1952 and 1967. On average, as we saw in \autoref{tab: summary}, unemployment rates are relatively low at less than 2\% throughout this period. However, there is substantial variation across areas within years, with some areas experiencing unemployment rates over 6\% and even 8\% in the early 1960s.

\begin{figure}[htp!]
\begin{center}
        \subfloat[Variation in 1951 industry shares    \label{fig:shares}]{    \includegraphics[trim=0 2cm 0 0,clip,width=0.79\linewidth]{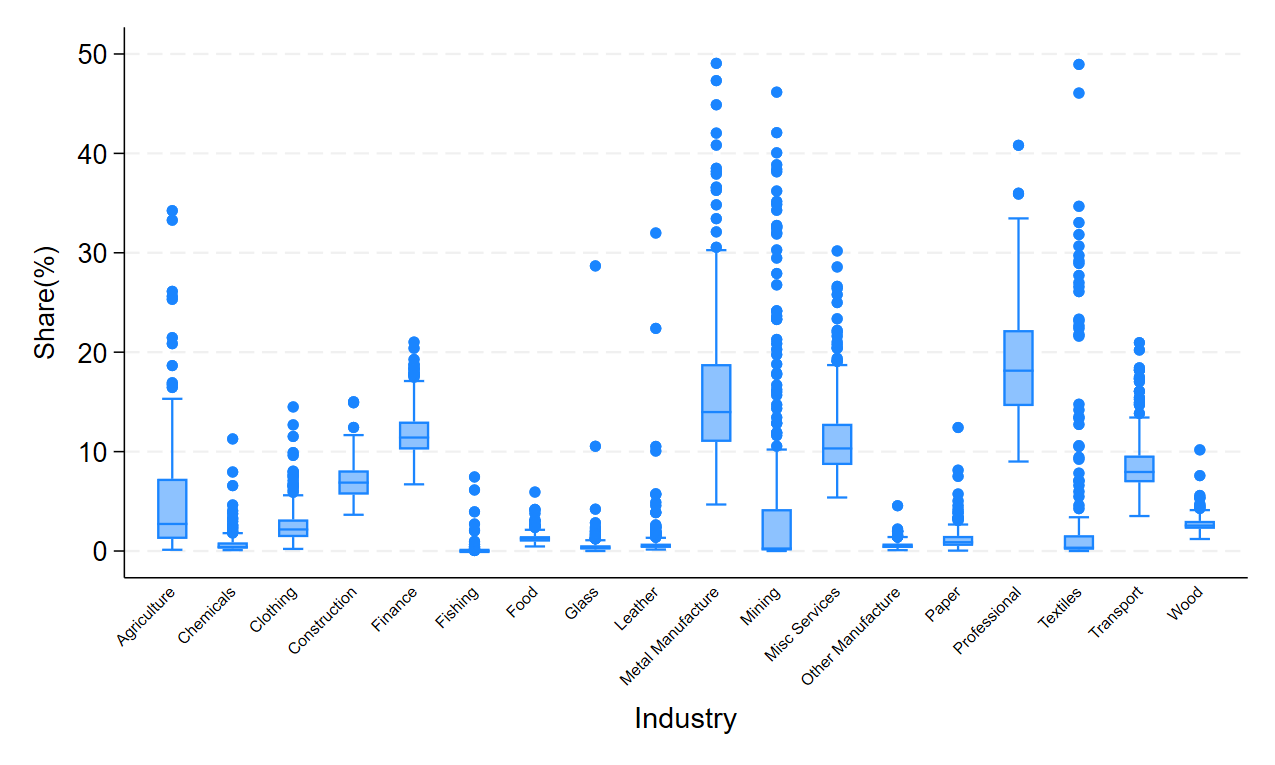}}\\
     \subfloat[Variation in national industry-specific unemployment    \label{fig:shocks}]{ \includegraphics[trim=0 2cm 0 0,clip,width=0.79\linewidth]{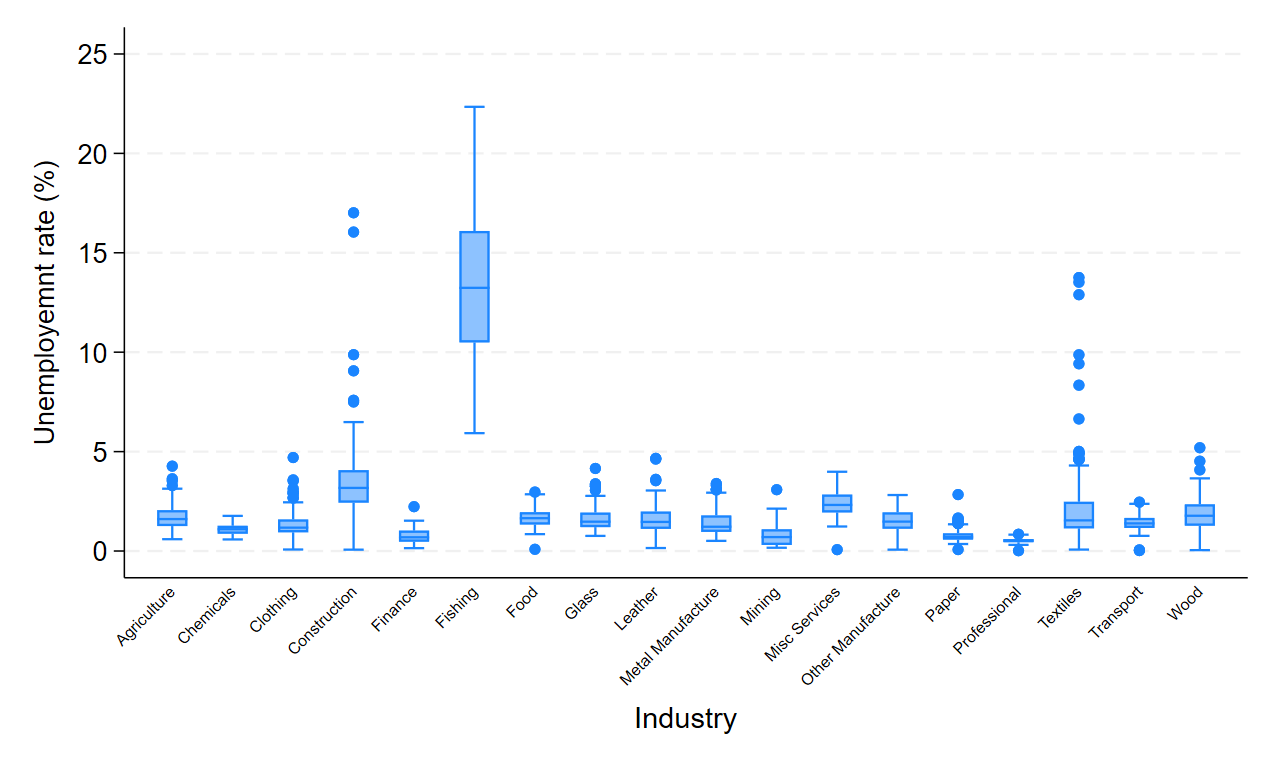}}
    \caption{Variation in local industry shares and national industry-specific unemployment}
        \label{fig:shares and shocks}
         \begin{minipage}{0.9\linewidth}
    \hfill \vspace{6pt}\\
     \footnotesize{Notes: Subfigure (a) plots the area-level share employed in each industry. Subfigure (b) plots the monthly unemployment rate between 1952 and 1967 by industry. The horizontal line inside the box shows the median, the lower line in the box is the 25th percentile and the upper line is the 75th percentile. The whiskers extend from the box by 1.5 times the inter-quartile range or to the minimum / maximum value (if this is nearer), the dots show any values beyond the whiskers.}
    \end{minipage}
\end{center}
 
\end{figure}

We instrument these area-level unemployment rates using the product of the 1951 area-level industry shares and the national industry-specific unemployment rates. The variation in the two components of our instrument is presented in \autoref{fig:shares and shocks}. Figure~\ref{fig:shares} shows substantial variation in industry shares across areas in 1951 (e.g. Metal Manufacturing, Mining and Textiles each account for almost 50\% of employment in at least one area and less than 10\% in others), with few industries showing relatively constant shares across areas (e.g., Food, Other Manufacturing and Wood). Figure~\ref{fig:shocks} shows the variation in national industry-level monthly unemployment rates between 1952 and 1967. It shows that some industries experienced large temporal variation in their national unemployment rates during this period (e.g., Fishing, Construction), whereas others experienced almost none (e.g., Chemicals, Professional).

\clearpage
\section{Results}\label{sec:results}
\subsection{OLS results}
We start by examining the conditional correlations between early-life exposure to adverse economic conditions, as proxied by the local unemployment rate, and later-life human capital and health. \autoref{tab: FE results} presents the estimates from separate OLS regressions, where we control for gender, month-of-birth, year-of-birth, and area-of-birth fixed effects as well as area-of-birth-specific linear trends. In specifications without fixed effects or area-specific trends (odd-numbered columns), we observe that higher unemployment in the first 1,000 days is associated with fewer years of schooling, lower wages, lower fluid intelligence, and shorter stature later in life. The coefficients are very similar for the three exposure periods. 

Once we control for the fixed effects as well as area-of-birth specific trends (even-numbered columns in \autoref{tab: FE results}) the associations are reduced substantially and no longer statistically different from zero. This highlights the importance of controlling for other spatial and temporal factors that could affect the causal link of interest. 

\bigskip 
\input{tables/main_fe_ols_results}

\FloatBarrier
\subsection{IV results}
To account for the potential endogeneity of local unemployment rates, as well as for the attenuation bias due to its likely measurement error, we turn to the instrumental variables specification. The first-stage results are reported in the odd-numbered columns of \autoref{tab: main iv results separately}, where each cell reports the estimate of interest from a separate regression. For example, the top left cell (Panel A, column 1) presents the coefficient for the impact of the weighted national unemployment rate during the pregnancy period on the area-level unemployment rate experienced during pregnancy for the sample of individuals for whom we observe their years of education. Hence, variation in the first-stage results across the columns in the same row is solely driven by differences in sample sizes across outcomes. 

The first-stage results show a strong relationship between the area-level unemployment rate and the instrumental variable. More specifically, a one percentage point increase in the instrument increases local unemployment by around 1.5 percentage points. Mechanically we would expect a one-to-one relationship, but local spill-over effects may explain why the coefficient is larger than one.\footnote{Indeed, when we exclude ``Finance'' and ``Professional'' industries from our instrument (\autoref{tab: main iv results no fi prof}, \autoref{sec:AppB}), the first-stage coefficient is close to one. This suggests the presence of spill-overs from shocks to these two industries, either due to their importance to the local economy or due to the services provided by these industries being closely linked with other (unobserved) industries.} Across all columns and panels of \autoref{tab: main iv results separately}, the association between our instrument and local unemployment is highly statistically significant, with first-stage F-statistics ranging between 131 and 200 depending on the period of unemployment (i.e., pregnancy, age 0 or age 1) and the outcome of interest. Note that if we specify the unemployment rate in the first stage as a function of each of the three period-specific instruments (i.e., including the weighted unemployment during pregnancy, age 0 as well as age 1), it clearly shows the relevance of our instrument and the importance of its timing. \autoref{tab: 1000 days first}, \autoref{sec:AppB}, reports these estimates, showing that the instrument for the pregnancy period is mainly predictive of unemployment during pregnancy and not during other times in early life. Similarly, the instrument measured at age 0 (age 1) is mainly predictive of unemployment at age 0 (age 1) and not at other ages. Thus, this presents evidence that our instrument is strong, but also that it is specific to the period it has been constructed to capture.

The IV estimates are presented in the even-numbered columns of \autoref{tab: main iv results separately}. As shown in columns (2), (4) and (8), we find small and statistically insignificant point estimates across all three exposure periods for educational attainment, wages and height, suggesting that these outcomes are unaffected by variations in local economic conditions early in life, whether during pregnancy or early childhood.

\input{tables/main_bartik_results}

For fluid intelligence, shown in column (6), we find a positive and marginally significant estimate in Panel B, suggesting that individuals who were exposed to adverse economic conditions during their first year of life have higher fluid intelligence scores. More specifically, a one percentage point increase in the average unemployment rate in the first year of life is estimated to increase fluid intelligence scores by 0.04 standard deviations. Since this effect is very small and only marginally significant, we are cautious not to overinterpret it. In summary, these analyses suggest that common, small fluctuations in local economic conditions during the early childhood period have little impact on individuals in older age. 

We replicate these analyses in \autoref{tab: 1000 days iv together}, \autoref{sec:AppB}, where instead of separately including unemployment rates in each of the three periods, we include all three unemployment variables simultaneously, thereby accounting for their mutual correlations. The estimates are largely similar to those in \autoref{tab: main iv results separately}, though with larger standard errors due to the correlation between unemployment rates across the three intervals (and with that, smaller first-stage F-statistics of around 50).

These analyses include only individuals born in one of the 218 (out of 310) areas for which we observe local monthly unemployment rates. However, these (more urban) areas are unlikely to be representative of all parts of England and Wales. We therefore explore the sensitivity of our results to this geographic selection. We do this by constructing the instrument at a lower-level geography (the district level), for which we observe both the predetermined shares as well as national industry-specific unemployment rates (but not the endogenous unemployment rates, which are only available at the area level). We then run reduced form regressions for the full sample covering 1,449 districts observed in the UK Biobank (out of a total of 1,470 in England and Wales). The results, presented in \autoref{tab: reduced form whole sample} of \autoref{sec:AppB}, are very similar to the reduced form results on the restricted (urban) sample (\autoref{tab: reduced form estimation sample} in \autoref{sec:AppB}) suggesting that the geographic selection of our estimation sample does not have a strong impact on our findings.

\subsection{Heterogeneity} 
Although we find no evidence that fluctuations in the local unemployment rate experienced during the first 1,000 days shape individuals' outcomes in the long run, this null effect may conceal heterogeneous effects across subgroups. To explore this, we investigate heterogeneity across three important dimensions: heterogeneity by genetic ``predisposition'', by gender and by socio-economic status (SES). To conduct these heterogeneity analyses, we merge our estimation sample with genetic data from the UK Biobank and area-level information on social classes from the 1951 Census.

\subsubsection{Heterogeneity by genetic variation} 
We start by examining heterogeneity with respect to individuals' genetic ``predisposition'' towards each of the four outcomes. Although we cannot (easily) change individuals' genetic variation, we \textit{can} change the environment. Hence, a better understanding of whether and how the environment interacts with individuals' genetic ``predisposition'' can help shape policy. Such gene-environment interplay is especially important because genetic variation is passed on from one generation to the next. This implies that any impacts on one generation are likely to be propagated to their offspring, potentially affecting intergenerational mobility. Furthermore, the estimates from these analyses inform us about potential complementarities between endowments and investments, with recent literature suggesting the use of genetic variation as proxies for the former \citep{Muslimova2020, biroli2025economics}. 

We use outcome-specific polygenic indices (PGIs) to empirically proxy individuals' genetic ``endowments''. We construct the PGIs by first running a set of genome-wide association studies (GWAS), one for each of our outcomes on UK Biobank participants who are not part of (nor related to individuals in) our analysis sample. We then use the GWAS summary statistics from these analyses to calculate outcome-specific PGIs for the individuals in our analysis sample.\footnote{We discuss the genetic data and the PGI construction in more detail in \autoref{sec:AppC}.}

To examine whether the impacts of early-life economic conditions differ by individuals' genetics, we split our sample by the median of the corresponding PGI, and separately estimate our IV specification in each subsample. This gives us a set of estimates for individuals with a high (above median) and low (below median) genetic ``endowment'' respectively. \autoref{tab: genetic bartik} reports the results, showing only little evidence of genetic heterogeneity in the effects. 

For height, our results indicate that individuals with a low PGI are 0.2cm taller if they were exposed to adverse economic conditions during the prenatal period (Panel A, column 7) and 0.2cm shorter when exposed during the first year of life (Panel B, column 7). The estimate in column 6 of Panel B suggests that the positive impact on fluid intelligence in \autoref{tab: main iv results separately} may be driven by a positive effect on high PGI individuals. 
Overall however, we find little systematic evidence of heterogeneity by one's genetic variation, with most estimates not statistically different from zero. Although there are some small and marginally significant estimates, we cannot interpret them as strong evidence of genetic heterogeneity.

\input{tables/genetic_groups_bartik}

\subsubsection{Heterogeneity by gender} 
We next explore whether the impact of local economic conditions varies between male and female participants in the UK Biobank. We estimate the impact using our IV specification for males and females respectively, with results reported in \autoref{tab: gender bartik}. Overall, we find little evidence of heterogeneity by gender, with most estimates being insignificantly different from zero and from each other.

\input{tables/gender_bartik}

\subsubsection{Heterogeneity by socio-economic status} 
Finally, we investigate heterogeneity in treatment effects by socio-economic status (SES). Since we do not observe participants' (or their parents') SES at birth, we proxy SES using the socio-economic composition of their area of birth as reported in the 1951 Census \citep{VisionOfBritain}. These data are again predetermined relative to our treatment of interest. The Census reports data on the proportion of each area's residents that are in social classes I to V.\footnote{The 1951 Census classifies individuals into five social classes based on occupation: (I) Professional, (II) Intermediate, (III) Skilled (manual and non-manual), (IV) Partly Skilled, and (V) Unskilled \citep{regoffice_1960}.} We use this to define an area as ``high SES'' when its combined share of residents in the three higher social classes (I, II and III) is above the median share across all areas, and ``low SES'' if it is below the median.

In \autoref{tab: SES bartik} we estimate our IV specification for individuals born in high and low SES areas respectively. Overall, we find no clear evidence of heterogeneity by socio-economic composition of the area of birth. While estimates in some cases differ substantially in magnitude, the standard errors are large, making most of them statistically indistinguishable from zero and from each other. 

\input{tables/SES_groups_bartik.tex}

\section{Selection}\label{sec:selection}
Our analysis consistently suggests that small, common fluctuations in local economic conditions experienced in the first 1,000 days of life do not affect individuals in older age. We explore potential reasons for our non-results, examining whether they are driven by the existence of selective participation in the UK Biobank, as well as fertility and mortality selection.

These analyses are motivated by the fact that local unemployment rates can affect household resources and parental employment, changing households' time and budget constraints. This may lead households to re-optimise their choices, potentially bringing forward or postponing any family planning decisions. In other words, it may be that the households having children during times of worse local economic conditions were systematically different from households having children during better times. For example, increased fertility of wealthy households during adverse periods may mask any negative effects on long-term outcomes. Furthermore, effects of adverse economic conditions on mortality (in childhood or in the long-run) as well as on long-term health can lead to selection issues in our sample.
We explore these potential channels in detail here.

We use data from \citet{Southall2004}, who digitised district-level statistics on demographics and infant mortality from the Registrar General Reports from 1930 to 1974 \citep{RegistrarGeneralReports}. Based on annual district-level birth data from these reports, we define the UK Biobank participation share as the ratio of the number of UK Biobank participants born in a given area and year to the total number of births in this area and year. Using annual total births and population counts, we construct the district-level birth rate (i.e., number of live births per 1,000 population) allowing us to investigate potential fertility selection. We additionally use information on infant mortality (i.e., the number of deaths in the first year of life per 1,000 live births) to examine potential mortality selection. As all of these data are reported at the district level, we aggregate these to the area level, similar to our employment and unemployment statistics.  

Unlike the unemployment data that is available at the monthly level, we observe fertility and mortality data annually. We therefore also collapse the unemployment data to the annual level, constructing the unemployment rate for each area-year by averaging each area's monthly unemployment rates across the year. We then use the same IV approach as in the main analysis to investigate the impact of local economic conditions on local fertility and mortality. To ensure comparability across samples, we use the same 218 areas and years (i.e., 1952-1967) as in the main analysis.  

\subsection{Selective participation} 
Our analysis sample consists of individuals who volunteered for participation in the UK Biobank when aged between 40 and 69 years. The sample is not representative of the general UK population or of their birth cohorts \citep{fry2017}. If adverse economic conditions during early childhood decrease the likelihood of being in our analysis sample, e.g. due to health, mortality or migration effects, this could impede the discovery of long-term effects in our analysis.

To explore this, we investigate the effect of local unemployment on the following year's, current year's and previous year's cohort participation share in the UK Biobank. A positive (negative) impact of local unemployment would suggest that more (fewer) children are selected into the UK Biobank when born around times of higher unemployment.

\input{tables/birth_death_rate_IV}

Estimates of the impact of economic conditions in early life on participation in the UKB are presented in column (1) of \autoref{tab: birth death rate IV}, with the different panels showing the effects for different exposure periods. For example, Panel A shows that a 1 percentage point increase in the unemployment rate reduces the share of local births in the following year which are included in the UK Biobank by 0.7 participants per 1,000 births, which is insignificantly different from zero. In sum, we find no evidence that early life economic conditions affect selection into our sample. 

\subsection{Fertility selection} 
One potential reason that we do not observe any impact of local economic conditions during early life on later-life outcomes is that this is masked by a selective fertility response to economic conditions. We thus investigate the effect of local unemployment on the local birth rate in the following, current and previous year. The estimates are presented in column (2) of \autoref{tab: birth death rate IV}. Panel A shows that a 1 percentage point increase in unemployment rate increases the local birth rate in the following year by 0.04 births per 1,000 population. None of the estimates of the fertility impact are significantly different from zero, providing no evidence that fertility selection impacts our findings.

\subsection{Mortality selection} 
Another reason for not observing a significant impact of economic conditions during the first 1,000 days of life on later-life outcomes could be selective mortality. For illustration, let us consider the scenario that worse economic conditions in utero lead to increases in stillbirths and neonatal mortality among the most vulnerable children. This selective mortality may mask negative effects on health or human capital. To assess this, we study the link between economic conditions around childbirth and infant mortality rates, with the estimates shown in columns (3)-(6) of \autoref{tab: birth death rate IV}.  
None of the estimates are statistically significant and all are small in magnitude, indicating that mortality selection is also unlikely to have affected our main analysis.

\FloatBarrier
\section{Conclusion}\label{sec:conclusion}
A growing body of literature shows that exposure to adverse early-life conditions may have life-long consequences. While most previous research focuses on health outcomes and large shocks, less is known about the long-term effects of more common and smaller shocks on both health and human capital. This study contributes to this topic by providing new evidence on the long-term impact of early-life economic conditions on individuals in later-life. Specifically, by combining historical unemployment statistics from the \textit{Labour Gazette}, the Registrar General Reports, and the 1951 Census with recent data on UK Biobank participants, we investigate this in the context of individuals born in urban England and Wales between 1952 and 1967—a period characterized by relatively stable economic conditions with low unemployment rates. More specifically, we use the local unemployment rate that individuals were exposed to in early life as a proxy for economic conditions and investigate the effect on their later-life education, hourly wage, fluid intelligence and height. To address potential concerns regarding the endogeneity of local unemployment rates, we implement an instrumental variable approach, using the product of predetermined industry shares and national industry-specific unemployment rates as instruments. This allows us to isolate the causal effects of early-life economic conditions. 

Our results indicate that small fluctuations in local economic conditions during the prenatal and early childhood periods have no impact on later-life health or human capital development. We also find no evidence that these estimates conceal heterogeneity between different subgroups, including gender, socio-economic composition of the area, and genetic ``predisposition'' to the outcomes of interest. To better understand these null results, we study potential selection mechanisms, where we examine the impact of local unemployment rates on participation in the UK Biobank, fertility and mortality rates. Our results show no strong evidence that such selections affect our findings, further reinforcing the robustness of our conclusions. 

While previous studies highlighted long-term effects of large economic downturns, our research suggests that smaller economic fluctuations do not have such pronounced effects. These findings have important policy implications, suggesting that policymakers can allocate resources more efficiently by focusing on major economic disruptions.

\clearpage
\bibliographystyle{aea}  
\typeout{}\bibliography{main}
\addcontentsline{toc}{section}{References}

\clearpage
\begin{appendices}
\setcounter{figure}{0}
\setcounter{table}{0}
\renewcommand{\thetable}{A.\arabic{table}}
\renewcommand{\thefigure}{A.\arabic{figure}}

\section{Industries in the \textit{Labour Gazette} and the 1951 UK Census}\label{sec:AppA}
The \textit{Labour Gazette} and the 1951 UK Census both group industries by `type'. To construct our instrument, we match the industry types from the two datasets. \autoref{tab:Matched industry names}, column 1, shows the industry names specified in the 1951 Census. We match these to the industries in the \textit{Labour Gazette}, presented in column 2. Finally, column 3 lists the names that we use in our analyses. Note that the matched data excludes nine industries: Administrators and managers, Warehouse-men and storekeepers, Stationary engine drivers, Workers in unskilled occupations, Defence service and Other undefined workers from the 1951 Census. Similarly, ex-service personnel not in industry, Utilities, and Public administration within the \textit{Labour Gazette} did not have a respective match within the 1951 Census.

\begin{landscape}
\input{tables/Matched_industry_names.tex}
\end{landscape}

\setcounter{figure}{0}
\setcounter{table}{0}
\renewcommand{\thetable}{B.\arabic{table}}
\renewcommand{\thefigure}{B.\arabic{figure}}

\section{Additional tables and figures}\label{sec:AppB}

\input{tables/main_bartik_results_no_finance_professional}

\input{tables/1000_days_first}
\input{tables/1000_days_bartik}
\input{tables/Reduced_form_estimated.tex}
\input{tables/Reduced_form_whole_sample}

\clearpage
\setcounter{figure}{0}
\setcounter{table}{0}
\renewcommand{\thetable}{C.\arabic{table}}
\renewcommand{\thefigure}{C.\arabic{figure}}

\FloatBarrier
\section{Genetics}\label{sec:AppC}
Every human cell, except for sex cells, contains 46 chromosomes arranged in pairs, with each pair holding one copy inherited from the mother and one from the father. These chromosomes are made of double-stranded deoxyribo nucleic acid (DNA) which consists of a vast number of base pairs arranged in a sequence, forming the human genome. Variations in these base pairs at specific locations within the population are known as Single Nucleotide Polymorphisms (SNPs), and are one of the most frequently studied types of genetic variation. When two possible base pairs exist at a given location (i.e., two alleles), the more common base pair is called the major allele, while the less common one is called the minor allele. Since humans have two copies of each chromosome, an individual can have zero, one, or two copies of the minor allele at a specific location.

To identify SNPs that correlate with outcomes of interest, Genome-Wide Association Studies (GWASes) systematically test associations between outcomes and SNPs. Typically, GWAS samples have far more SNPs than individuals, making multivariate regression models impractical for identifying SNP effects jointly. To circumvent this issue, GWASes instead perform univariate regressions of the outcome on each SNP. Research has demonstrated that many social science outcomes are polygenic, that is, they are influenced by many SNPs, each contributing only a small effect. Consequently, to increase the predictive power of SNPs with respect to an outcome, it is common practice to combine the associations of many individual SNPs with the outcome into polygenic indices (PGIs). These indices, \( G_i \), are linear combinations of SNPs, with weights (\( \beta_j \)) obtained from an independent (i.e., non-overlapping) GWAS:
\[ G_i = \sum_{j} \beta_j X_{ij} \]
where \( X_{ij} \) represents the minor allele count (0, 1, or 2) at SNP \( j \) for individual \( i \). This approach follows from an additive genetic model, where each SNP adds to an individual's overall genetic predisposition \citep[see e.g.,][]{purcell2009prs}.

We run a GWAS for each of our outcomes to estimate the SNP weights. To avoid overfitting, we run our GWASes on a hold-out sample of UK Biobank individuals that are not in our analysis sample (see Section~\ref{sec:data} for the construction of the analysis sample). \autoref{tab:gwas-descriptives} reports the descriptives for the GWAS samples. We further prepare and quality control our genetic data as described in \citet{mitchell2019gwas} and use BOLT-LMM \citep{boltlmm} to run the GWASs on the subset of individuals that have European genetic ancestry. Following the existing literature, our GWASs control for genotyping array, sex and the first 10 genetic principal components as well as third-order polynomials in birth year by sex, allowing cohort trends to differ by sex.

Having estimated the SNP effects, we boost the predictive power of our polygenic indices by using LDpred2 \citep{prive2020ldpred2} to adjust the raw SNP weights to account for linkage disequilibrium (that is, correlations between neighbouring SNPs). To apply LDpred2, we assume an infinitesimal genetic model, and restrict our attention to the 1.6 million SNPs in Hapmap3 \citep{hapmap3} with a minor allele frequency larger than $0.01$ and info score (measure of imputation quality) above $0.97$. We standardise all polygenic indices to have zero mean and unit variance in the analysis sample.

To examine the predictive power of the polygenic indices, we regress each outcome onto its corresponding PGI as well as the first 10 genetic principal components and an indicator for gender. Table~\ref{tab:pgs_predictive} reports the estimates together with the incremental R$^2$, defined as the increase in R$^2$ when the polygenic index is included as a covariate. The results show that the polygenic indices strongly predict their target outcomes, with incremental R$^2$ ranging from 2.7\% to 16.8\%.

\input{tables/gwas_descriptives}
\input{tables/pgs_predictive_power.tex}

\end{appendices}
\end{document}

%% file: tables/summary_stat.tex
\begin{table}[htbp]\centering
\def\sym#1{\ifmmode^{#1}\else\(^{#1}\)\fi}
\caption{Summary statistics\label{tab: summary}}
\begin{threeparttable}
\begin{tabular}{l*{3}{cccc}}
\hline\hline
	  &\multicolumn{2}{c}{(1)}&\multicolumn{2}{c}{(2)}   \\
	  &\multicolumn{2}{c}{UK Biobank sample}&\multicolumn{2}{c}{Estimation sample}   \\ 
          &     Mean&       SD&     Mean&       SD\\
\hline
Male      &     0.46&     0.50&     0.45&     0.50\\
Years of education&    13.11&     2.33&    13.34&     2.20\\
Hourly wages&    14.66&     5.24&    14.77&     5.10\\
Fluid intelligence&     6.08&     2.14&     6.27&     2.13\\
Height (cm)&   168.44&     9.28&   169.55&     9.25\\
Area unemployment rate (\%)&    &    &     1.58&     1.16\\
Area instrument (\%)&     &     &     1.37&     0.39\\
\hline
Max observations  & \multicolumn{2}{c}{502,385}&\multicolumn{2}{c}{128,111} \\
\hline\hline
\end{tabular}
\begin{tablenotes}[para,flushleft]
     \footnotesize{ Notes: The summary statistics for gender 
     and the four outcomes of interest are at the individual level; the unemployment rate and the instrument are measured at the area level. }
     \footnotesize {The instrument is the product of predetermined area-level industry shares and national industry-specific unemployment rates. The number of observations varies depending on the variable considered.}
\end{tablenotes}
\end{threeparttable}
\end{table}

%% file: tables/main_fe_ols_results.tex
\begin{table}[htbp]\centering
\def\sym#1{\ifmmode^{#1}\else\(^{#1}\)\fi}
\caption{OLS estimates of the relationship between early life economic conditions and human capital / health outcomes \label{tab: FE results}}
\begin{adjustbox}{max width=\textwidth}
\begin{threeparttable}
\begin{tabularx}{1.1\linewidth}{Xcccccccc}
\hline\hline
               
                &\multicolumn{2}{c}{\makecell{Education \\ (years)}}&\multicolumn{2}{c}{\makecell{Log \\ hourly wages}}&\multicolumn{2}{c}{\makecell{Fluid \\ intelligence}}&\multicolumn{2}{c}{\makecell{Height \\ (cm)}}\\
                 &\multicolumn{1}{c}{(1)}&\multicolumn{1}{c}{(2)}&\multicolumn{1}{c}{(3)}&\multicolumn{1}{c}{(4)}&\multicolumn{1}{c}{(5)}&\multicolumn{1}{c}{(6)}&\multicolumn{1}{c}{(7)}&\multicolumn{1}{c}{(8)}\\
\hline
\multicolumn{8}{l}{\emph{Panel A: Local unemployment during pregnancy}}\\
$UE_{itz}^{age=-1}$  & -0.071** &0.010    & -0.008* &0.001   & -0.032*** &-0.008    &  -0.180***&0.050   \\
                             & (0.031)  &(0.015)  & (0.004) &(0.003) & (0.012)   &(0.011)   &  (0.061) &  (0.040)  \\[4pt]
\multicolumn{8}{l}{\emph{Panel B: Local unemployment at age 0}}\\
$UE_{itz}^{age=0}$  & -0.068** &-0.000 & -0.008* &0.001   & -0.035*** &-0.003    &  -0.160***&-0.004   \\
                             & (0.031)  &(0.013)  & (0.004) &(0.002) & (0.011)   &(0.010)   &  (0.060)  & (0.046)   \\[4pt]
\multicolumn{8}{l}{\emph{Panel C: Local unemployment at age 1}}\\
$UE_{itz}^{age=1}$  & -0.063** &0.004    & -0.007* &0.003   & -0.036*** &-0.0001   &  -0.132**& -0.014   \\
                             & (0.031)  &(0.017)  & (0.004) &(0.003) & (0.011)   &(0.009)   &  (0.058)   &(0.046)  \\ 
                           \hline
Observations                 & 109,858  &109,842  & 97,420  &97,406  & 54,614    &54,597    &  110,740 & 110,725   \\
\hline 
Fixed effects &No&Yes&No&Yes&No&Yes&No&Yes\\
\hline 
\hline 

\end{tabularx}
\begin{tablenotes}[para,flushleft]
     \footnotesize{Notes: Education is measured in years. Wages are measured as log hourly wages imputed using individuals' occupation. Fluid intelligence is standardised within the estimation sample to have mean 0, standard deviation 1. Height is individuals' standing height (in cm). We include controls for gender and month-of-birth fixed effects in all specifications. The even numbered columns also include year-of-birth fixed effects, area-of-birth fixed effects, and area-specific trends. Standard errors in parentheses are clustered at the area level. \sym{*} \(p<0.1\), \sym{**} \(p<0.05\), \sym{***} \(p<0.01\)}
\end{tablenotes}
\end{threeparttable}
\end{adjustbox}
\end{table}

%% file: tables/main_bartik_results.tex
\begin{table}[tbp]\centering
\def\sym#1{\ifmmode^{#1}\else\(^{#1}\)\fi}
\caption{IV estimates of the effect of early life economic conditions on individuals' human capital and health outcomes \label{tab: main iv results separately}}
\begin{adjustbox}{max width=\textwidth}
\begin{threeparttable}
\begin{tabularx}{1.1\linewidth}{Xcccccccc}
\hline\hline
           &\multicolumn{2}{c}{\makecell{Education \\ (years)}}&\multicolumn{2}{c}{\makecell{Log \\ hourly wages}}&\multicolumn{2}{c}{\makecell{Fluid \\ intelligence}}&\multicolumn{2}{c}{\makecell{Height \\ (cm)}}\\
                 &\multicolumn{1}{c}{(1)}&\multicolumn{1}{c}{(2)}&\multicolumn{1}{c}{(3)}&\multicolumn{1}{c}{(4)}&\multicolumn{1}{c}{(5)}&\multicolumn{1}{c}{(6)}&\multicolumn{1}{c}{(7)}&\multicolumn{1}{c}{(8)}\\
           &FS&IV&FS&IV&FS&IV&FS&IV\\
           \cline{2-9} \\[-0.5em]
           &$UE_{itz}^{age=a}$ &$Y_{itz}$&$UE_{itz}^{age=a}$ &$Y_{itz}$&$UE_{itz}^{age=a}$ &$Y_{itz}$&$UE_{itz}^{age=a}$ &$Y_{itz}$\\
           \\[-0.5em]
\hline
\multicolumn{5}{l}{\emph{Panel A: Local unemployment during pregnancy}}\\
$Z_{itz}^{age=-1}$    &    1.504***&   &    1.501***&  &    1.569***&  &    1.504***&   \\
                     &  (0.131)   &   &  (0.131)   &  &  (0.135)   &  &  (0.131)   &  \\

$UE_{itz}^{age=-1}$   &  &   -0.016   &   &   -0.004   &   &   -0.022   &  &    0.145   \\
          	     &  &  (0.043)   &   &  (0.006)   &   &  (0.026)   &  &  (0.108)   \\

F statistic 	     &   132.54 &  &   131.34 &  &   136.09  &  &   132.41 &  \\ [6pt]
\multicolumn{5}{l}{\emph{Panel B: Local unemployment at age 0}}\\
$Z_{itz}^{age=0}$    & 1.511***&  &    1.504***&   &    1.562***&    &    1.511***&     \\
                     &  (0.115)   & &  (0.114)   &   &  (0.122)   &     &  (0.115)   &      \\

$UE_{itz}^{age=0}$   &  &   -0.007   &  &    0.002   &   &    0.044*  &   &   -0.106   \\
                     &  &  (0.043)   &  &  (0.006)   &   &  (0.023)   &   &  (0.102)   \\
F statistic          &   172.34   &    &   173.51   &   &   163.77   &  &   173.84   &    \\
 [6pt]
\multicolumn{5}{l}{\emph{Panel C: Local unemployment at age 1}}\\
$Z_{itz}^{age=1}$   &   1.492***&   &    1.493***&  &    1.569***&  &    1.494***&     \\
                    &  (0.107)  &   &  (0.105)   &  &  (0.119)   &  &  (0.108)   &      \\

$UE_{itz}^{age=1}$  &  &   -0.008   &  &   -0.003   &  &    0.006   &  &   -0.101   \\
                    &  &  (0.038)   &  &  (0.006)   &  &  (0.020)   &  &  (0.108)   \\
F statistic         &   192.98   &  &   200.32   &  &   174.62   &  &   191.47   &   \\

\hline
Observations       & 109,842  & 109,842& 97,406     & 97,406  & 54,597    & 54,597   & 110,725 & 110,725   \\

\hline\hline

\end{tabularx}
\begin{tablenotes}[para,flushleft]
     \footnotesize{Notes: Education is measured in years. Wages are measured as log hourly wages imputed using individuals' occupation. Fluid intelligence is standardised within the estimation sample to have mean 0, standard deviation 1. Height is individuals' standing height (in cm). We include controls for gender, month-of-birth and year-of-birth fixed effects, area-of-birth fixed effects, and area-specific trends. Standard errors in parentheses are clustered at area level. \sym{*} \(p<0.1\), \sym{**} \(p<0.05\), \sym{***} \(p<0.01\)}
\end{tablenotes}
\end{threeparttable}
\end{adjustbox}
\end{table}

%% file: tables/genetic_groups_bartik.tex
\begin{table}[htbp]\centering
\def\sym#1{\ifmmode^{#1}\else\(^{#1}\)\fi}
\caption{IV estimates of the effect of early life economic conditions --- Genetic heterogeneity \label{tab: genetic bartik}}
\begin{adjustbox}{max width=\textwidth}
\begin{threeparttable}
\begin{tabular}{l*{8}{c}}
\hline\hline
		&\multicolumn{2}{c}{Education (years)}&\multicolumn{2}{c}{Log hourly wage}&\multicolumn{2}{c}{Fluid intelligence}&\multicolumn{2}{c}{Height (cm)} \\
          	&\multicolumn{1}{c}{(1)}   &\multicolumn{1}{c}{(2)}   &\multicolumn{1}{c}{(3)}   &\multicolumn{1}{c}{(4)}   &\multicolumn{1}{c}{(5)}   &\multicolumn{1}{c}{(6)}   &\multicolumn{1}{c}{(7)}   &\multicolumn{1}{c}{(8)}   \\
		&\multicolumn{1}{c}{Low PGI}&\multicolumn{1}{c}{High PGI}&\multicolumn{1}{c}{Low PGI}&\multicolumn{1}{c}{High PGI}&\multicolumn{1}{c}{Low PGI}&\multicolumn{1}{c}{High PGI}&\multicolumn{1}{c}{Low PGI}&\multicolumn{1}{c}{High PGI}  \\
\hline
\multicolumn{8}{l}{\emph{Panel A: Local unemployment during pregnancy}}\\
$UE_{itz}^{age=-1}$ &    0.029   &   -0.038   &    0.002   &   -0.006   &    0.006   &   -0.065   &    0.210*  &    0.082   \\
          &  (0.055)   &  (0.052)   &  (0.008)   &  (0.008)   &  (0.035)   &  (0.041)   &  (0.123)   &  (0.146)   \\
F statistic&   127.93   &   132.33   &   126.45   &   126.08   &   124.61   &   140.58   &   130.22   &   129.88   \\ [3pt]

\multicolumn{8}{l}{\emph{Panel B: Local unemployment at age 0}}\\
$UE_{itz}^{age=0}$ &   -0.001   &   -0.026   &    0.006   &   -0.001   &    0.017   &    0.058*  &   -0.238*  &    0.044   \\
          &  (0.066)   &  (0.063)   &  (0.008)   &  (0.009)   &  (0.030)   &  (0.032)   &  (0.142)   &  (0.139)   \\

F statistic&   156.91   &   184.68   &   147.81   &   198.68   &   147.29   &   169.68   &   174.59   &   168.16   \\ [3pt]

\multicolumn{8}{l}{\emph{Panel C: Local unemployment at age 1}}\\
$UE_{itz}^{age=1}$ &   -0.040   &    0.008   &   -0.006   &   -0.0004   &   -0.022   &    0.037   &    0.151   &   -0.259   \\
          &  (0.062)   &  (0.049)   &  (0.009)   &  (0.011)   &  (0.031)   &  (0.034)   &  (0.130)   &  (0.157)   \\
F statistic&   181.24   &   189.36   &   185.74   &   194.09   &   160.40   &   176.88   &   204.48   &   165.24   \\
\hline

Observations&   51,902   &   52,114   &   45,194   &   46,822   &   25,156   &   25,878   &   52,177   &   52,192   \\

\hline\hline
\end{tabular}
\begin{tablenotes}[para,flushleft]
     \footnotesize{Notes: Education is measured in years. Wages are measured as log hourly wages imputed using individuals' occupation. Fluid intelligence is standardised within the estimation sample to have mean 0, standard deviation 1. Height is individuals' standing height (in cm). We add controls for gender, month-of-birth and year-of-birth fixed effects, area-of-birth fixed effects, and area-specific trends. We include the first 10 genetic principal components in the regressions. We define an individual's PGI as `high PGI’ if the PGI is higher than or equal to the median of the corresponding polygenic index, otherwise it is `low PGI'. Standard errors in parentheses are clustered at area level. \sym{*} \(p<0.1\), \sym{**} \(p<0.05\), \sym{***} \(p<0.01\)}
\end{tablenotes}
\end{threeparttable}
\end{adjustbox}
\end{table}

%% file: tables/gender_bartik.tex
\begin{table}[htbp]\centering
\def\sym#1{\ifmmode^{#1}\else\(^{#1}\)\fi}
\caption{IV estimates of the effect of early life economic conditions --- Gender heterogeneity \label{tab: gender bartik}}
\begin{adjustbox}{max width=\textwidth}
\begin{threeparttable}
\begin{tabular}{l*{8}{c}}
\hline\hline

		&\multicolumn{2}{c}{Education (years)}&\multicolumn{2}{c}{Log hourly wage}&\multicolumn{2}{c}{Fluid intelligence}&\multicolumn{2}{c}{Height (cm)} \\
          	&\multicolumn{1}{c}{(1)}   &\multicolumn{1}{c}{(2)}   &\multicolumn{1}{c}{(3)}   &\multicolumn{1}{c}{(4)}   &\multicolumn{1}{c}{(5)}   &\multicolumn{1}{c}{(6)}   &\multicolumn{1}{c}{(7)}   &\multicolumn{1}{c}{(8)}   \\
		&\multicolumn{1}{c}{Male}&\multicolumn{1}{c}{Female}&\multicolumn{1}{c}{Male}&\multicolumn{1}{c}{Female}&\multicolumn{1}{c}{Male}&\multicolumn{1}{c}{Female}&\multicolumn{1}{c}{Male}&\multicolumn{1}{c}{Female}\\
\hline
\multicolumn{8}{l}{\emph{Panel A: Local unemployment during pregnancy}}\\
$UE_{itz}^{age=-1}$ &   -0.113** &    0.064   &   -0.010   &    0.001   &   -0.017   &   -0.027   &    0.303*  &    0.018   \\
          &  (0.053)   &  (0.064)   &  (0.008)   &  (0.007)   &  (0.039)   &  (0.032)   &  (0.174)   &  (0.124)   \\
F statistic&   124.68   &   138.33   &   126.24   &   134.89   &   130.84   &   136.36   &   125.58   &   137.73   \\ [5pt]
\multicolumn{8}{l}{\emph{Panel B: Local unemployment at age 0}}\\
$UE_{itz}^{age=0}$ &   -0.033   &    0.012   &    0.001   &    0.003   &    0.043   &    0.043   &   -0.143   &   -0.070   \\
          &  (0.058)   &  (0.055)   &  (0.009)   &  (0.009)   &  (0.031)   &  (0.031)   &  (0.175)   &  (0.153)   \\
F statistic&   159.69   &   178.49   &   164.75   &   175.18   &   159.68   &   158.21   &   160.78   &   180.20   \\ [5pt]

\multicolumn{8}{l}{\emph{Panel C: Local unemployment at age 1}}\\
$UE_{itz}^{age=1}$ &    0.018   &   -0.028   &    0.004   &   -0.009   &    0.030   &   -0.009   &   -0.151   &   -0.069   \\
          &  (0.059)   &  (0.052)   &  (0.009)   &  (0.007)   &  (0.035)   &  (0.025)   &  (0.216)   &  (0.114)   \\
F statistic&   229.60   &   166.43   &   234.15   &   173.99   &   215.34   &   143.28   &   224.55   &   166.75   \\
\hline
Observations&   48,798   &   61,026   &   44,180   &   53,207   &   24,121   &   30,455   &   49,185   &   61,522   \\

\hline\hline

\end{tabular}
\begin{tablenotes}[para,flushleft]
     \footnotesize{Notes: Education is measured in years. Wages are measured as log hourly wages imputed using individuals' occupation. Fluid intelligence is standardised within the estimation sample to have mean 0, standard deviation 1. Height is individuals' standing height (in cm). We include controls for month-of-birth and year-of-birth fixed effects, area-of-birth fixed effects, and area-specific trends. Standard errors in parentheses are clustered at area level. \sym{*} \(p<0.1\), \sym{**} \(p<0.05\), \sym{***} \(p<0.01\)}
\end{tablenotes}
\end{threeparttable}
\end{adjustbox}
\end{table}

%% file: tables/SES_groups_bartik.tex
\begin{table}[htbp]\centering
\def\sym#1{\ifmmode^{#1}\else\(^{#1}\)\fi}
\caption{IV estimates of the effect of early life economic conditions --- Heterogeneity by SES \label{tab: SES bartik}}
\begin{adjustbox}{max width=\textwidth}
\begin{threeparttable}
\begin{tabular}{l*{8}{c}}
\hline\hline
		&\multicolumn{2}{c}{Education (years)}&\multicolumn{2}{c}{Log hourly wage}&\multicolumn{2}{c}{Fluid intelligence}&\multicolumn{2}{c}{Height (cm)} \\
          	&\multicolumn{1}{c}{(1)}   &\multicolumn{1}{c}{(2)}   &\multicolumn{1}{c}{(3)}   &\multicolumn{1}{c}{(4)}   &\multicolumn{1}{c}{(5)}   &\multicolumn{1}{c}{(6)}   &\multicolumn{1}{c}{(7)}   &\multicolumn{1}{c}{(8)}   \\
		&\multicolumn{1}{c}{Low SES}&\multicolumn{1}{c}{High SES}&\multicolumn{1}{c}{Low SES}&\multicolumn{1}{c}{High SES}&\multicolumn{1}{c}{Low SES}&\multicolumn{1}{c}{High SES}&\multicolumn{1}{c}{Low SES}&\multicolumn{1}{c}{High SES}  \\
\hline
\multicolumn{8}{l}{\emph{Panel A: Local unemployment during pregnancy}}\\
$UE_{itz}^{age=-1}$&   -0.037   &    0.007   &   -0.006   &   -0.002   &   -0.013   &   -0.035   &    0.223*  &    0.060   \\
          &  (0.049)   &  (0.083)   &  (0.006)   &  (0.011)   &  (0.036)   &  (0.045)   &  (0.118)   &  (0.203)   \\
F statistic&    66.52   &   181.41   &    64.09   &   187.93   &    87.32   &   127.43   &    66.15   &   182.92   \\ [3pt]
\multicolumn{8}{l}{\emph{Panel B: Local unemployment at age 0}}\\
$UE_{itz}^{age=0}$ &   -0.007   &   -0.017   &   -0.003   &    0.008   &    0.010   &    0.093** &   -0.087   &   -0.175   \\
          &  (0.061)   &  (0.064)   &  (0.007)   &  (0.010)   &  (0.023)   &  (0.045)   &  (0.129)   &  (0.180)   \\
F statistic&    83.14   &   253.47   &    81.47   &   254.59   &   118.47   &   219.49   &    83.54   &   250.97   \\ [3pt]
\multicolumn{8}{l}{\emph{Panel C: Local unemployment at age 1}}\\
$UE_{itz}^{age=1}$ &    0.019   &   -0.044   &   -0.000   &   -0.006   &   -0.020   &    0.035   &   -0.273*  &    0.027   \\
          &  (0.059)   &  (0.055)   &  (0.007)   &  (0.009)   &  (0.028)   &  (0.035)   &  (0.152)   &  (0.174)   \\
F statistic&    83.05   &   194.68   &    88.56   &   198.00   &    97.82   &   134.56   &    82.31   &   193.85   \\
\hline
Observations&   43,902   &   65,940   &   38,452   &   58,954   &   23,519   &   31,078   &   44,107   &   66,618   \\

\hline\hline
\end{tabular}
\begin{tablenotes}[para,flushleft]
     \footnotesize{Notes: Education is measured in years. Wages are measured as log hourly wages imputed using individuals' occupation. Fluid intelligence is standardised within the estimation sample to have mean 0, standard deviation 1. Height is individuals' standing height (in cm). We include controls for gender, month-of-birth and year-of-birth fixed effects, area-of-birth fixed effects, and area-specific trends. We define an area as `high SES' when its combined share of social classes I, II and III is above the median share across all areas, otherwise it is `low SES'. Standard errors in parentheses are clustered at area level. \sym{*} \(p<0.1\), \sym{**} \(p<0.05\), \sym{***} \(p<0.01\)}
\end{tablenotes}
\end{threeparttable}
\end{adjustbox}
\end{table}

%% file: tables/birth_death_rate_IV.tex
\begin{table}[htbp]\centering
\def\sym#1{\ifmmode^{#1}\else\(^{#1}\)\fi}
\caption{IV estimates of the effect of early life economic conditions on participation in the UK Biobank, fertility and mortality \label{tab: birth death rate IV}}
\begin{threeparttable}
\begin{tabular}{l*{6}{c}}
\hline\hline
                              &\multicolumn{1}{c}{(1)}&\multicolumn{1}{c}{(2)}&\multicolumn{1}{c}{(3)}&\multicolumn{1}{c}{(4)}&\multicolumn{1}{c}{(5)} &\multicolumn{1}{c}{(6)} \\
                &\multicolumn{1}{c}{\makecell{UKB \\ participation \\ share}}&\multicolumn{1}{c}{\makecell{Birth \\ rate}}&\multicolumn{1}{c}{\makecell{Death rate \\ $<$ 1 week}}&\multicolumn{1}{c}{\makecell{Death rate \\ $<$ 4 weeks}}&\multicolumn{1}{c}{\makecell{Death rate \\ $<$ 1 year}}&\multicolumn{1}{c}{\makecell{Stillbirth \\ rate}}\\
\hline
\multicolumn{7}{l}{\emph{Panel A: Local unemployment during previous year}}\\
 UE rate (\%)&   -0.706   &    0.044   &    0.753   &    0.253   &   -0.225   &    0.453   \\
          &  (1.063)   &  (0.099)   &  (1.031)   &  (0.608)   &  (0.785)   &  (0.866)   \\
F statistic&    61.42   &    61.42   &    36.64   &    60.93   &    60.61   &    61.20   \\ [6pt]

\multicolumn{7}{l}{\emph{Panel B: Local unemployment during current year}}\\
UE rate (\%)&   -0.629   &   -0.052   &    0.430   &    0.670   &   -0.273   &   -0.127   \\
          &  (0.591)   &  (0.159)   &  (1.280)   &  (0.692)   &  (0.820)   &  (1.092)   \\
F statistic&    45.87   &    45.87   &    23.84   &    47.08   &    45.92   &    45.96   \\ [6pt]

\multicolumn{7}{l}{\emph{Panel C: Local unemployment during following year}} \\
UE rate (\%)&    1.217   &    0.078   &   -0.392   &   -0.714   &   -0.401   &    1.270   \\
          &  (0.756)   &  (0.157)   &  (1.721)   &  (0.643)   &  (0.860)   &  (0.800)   \\
F statistic&    53.71   &    53.71   &    24.48   &    53.37   &    53.53   &    53.67   \\

\hline 
Observations&    1,824   &    1,824   &    1,273   &    1,819   &    1,823   &    1,823   \\

\hline\hline

\end{tabular}
\begin{tablenotes}[para,flushleft]
	\footnotesize{Notes: The UKB participation share is the number of UKB participants per 1,000 births in the same area and year of birth, the birth rate is the area's annual number of births per 1,000 population, the death rates are an area's annual number of deaths (in the given age periods) per 1,000 births, the stillbirth rate is the area's annual number of stillbirths per 1,000 births. We include controls for year fixed effects, area fixed effects, and area-specific linear trends. Standard errors in parentheses are clustered at area level. \sym{*} \(p<0.1\), \sym{**} \(p<0.05\), \sym{***} \(p<0.01\)}
\end{tablenotes}
\end{threeparttable}
\end{table}

%% file: tables/Matched_industry_names.tex
\begin{table}[htbp] \centering
\def\sym#1{\ifmmode^{#1}\else\(^{#1}\)\fi}
\caption{Industry mapping between 1951 UK Census and Labour Gazette \label{tab:Matched industry names}}
\begin{threeparttable}
\begin{tabular}{|l|l|l|}
\hline

\textbf{Industries from 1951 UK Census}                          & \textbf{Industries from Labour Gazette}  & \textbf{Matched industry names}             \\ \hline
Fishermen                                                        & Fishing                                  & Fishing                            		      \\ \hline
Agricultural, etc occupations                                    & Agriculture, horticulture and forestry   & Agriculture                        		      \\ \hline
Mining and quarrying occupations                                 & Mining and quarrying                     & Mining                            		      \\ \hline
Workers in ceramics, glass, cement, etc.                         & Bricks, pottery, glass and cement        & Glass                              		      \\ \hline
Coal gas, etc, makers, workers in chemicals                      & Chemicals and allied industries          & Chemicals                          		      \\ \hline
\multirow{5}{*}{Workers in metal manufacture, engineering}       & Metal manufacture                        & \multirow{5}{*}{Metal manufacturing} 		      \\ \cline{2-2}
                                                                 & Vehicles                                 &                                    		      \\ \cline{2-2}
                                                                 & Engineering                              &                                    		      \\ \cline{2-2}
                                                                 & Metal industries                         &                                    		      \\ \cline{2-2}
                                                                 & Shipbuilding and marine engineering      &                                    		      \\ \hline
Textile workers                                                  & Textiles                                 & Textiles                           		      \\ \hline
Leather workers, fur dressers                                    & Leather and fur                          & Leather                            		      \\ \hline
Makers of textile goods and articles of dress                    & Clothing                                 & Clothing                           		      \\ \hline
Makers of foods, drinks and tobacco                              & Food, drink and tobacco                  & Food                               		      \\ \hline
Workers in wood, cane and cork                                   & Wood manufacture                         & Wood                               		      \\ \hline
Makers of, workers in paper; printers                            & Paper and printing                       & Paper                              		      \\ \hline
Makers of products (n.e.s)                                       & Other manufacturing industries           & Other manufacturing                  		      \\ \hline
Workers in building and contracting                              & \multirow{2}{*}{Construction}            & \multirow{2}{*}{Construction}      		      \\ \cline{1-1}
Painters and decorators                                          &                                          &                                    		      \\ \hline
Professional and technical (exc. clerical)                       & \multirow{2}{*}{Professional services}   & \multirow{2}{*}{Professional}      		      \\ \cline{1-1}
Clerks, typists, etc                                             &                                          &                                    		      \\ \hline
\multirow{2}{*}{Persons employed in transport, etc.}             & Transport                                & \multirow{2}{*}{Transport}         		      \\ \cline{2-2}
                                                                 & Distributive trades                      &                                    		      \\ \hline
Commercial, finance, etc. (exc. clerical)                        & Insurance, banking and finance           & Finance                            		      \\ \hline
Persons engaged in entertainments and sport                      & \multirow{2}{*}{Misc services}           & \multirow{2}{*}{Misc services}     		      \\ \cline{1-1}
Persons engaged in personal service                              &                                          &                                    		      \\ \hline

\end{tabular}
\begin{tablenotes}[para,flushleft]
     \footnotesize{}
\end{tablenotes}
\end{threeparttable}

\end{table}

%% file: tables/main_bartik_results_no_finance_professional.tex
\begin{table}[htbp]\centering
\def\sym#1{\ifmmode^{#1}\else\(^{#1}\)\fi}
\caption{IV estimates excluding ``Finance'' and ``Professional'' industries \label{tab: main iv results no fi prof}}
\begin{adjustbox}{max width=\textwidth}
\begin{threeparttable}
\begin{tabularx}{1.1\linewidth}{Xcccccccc}
\hline\hline
           &\multicolumn{2}{c}{\makecell{Education \\ (years)}}&\multicolumn{2}{c}{\makecell{Log \\ hourly wages}}&\multicolumn{2}{c}{\makecell{Fluid \\ intelligence}}&\multicolumn{2}{c}{\makecell{Height \\ (cm)}}\\
                 &\multicolumn{1}{c}{(1)}&\multicolumn{1}{c}{(2)}&\multicolumn{1}{c}{(3)}&\multicolumn{1}{c}{(4)}&\multicolumn{1}{c}{(5)}&\multicolumn{1}{c}{(6)}&\multicolumn{1}{c}{(7)}&\multicolumn{1}{c}{(8)}\\
           &FS&IV&FS&IV&FS&IV&FS&IV\\
           &$UE_{itz}^{age=a}$ &$Y_{itz}$&$UE_{itz}^{age=a}$ &$Y_{itz}$&$UE_{itz}^{age=a}$ &$Y_{itz}$&$UE_{itz}^{age=a}$ &$Y_{itz}$\\
           \\[-0.5em]
\hline
\multicolumn{5}{l}{\emph{Panel A: Local unemployment during pregnancy}}\\
$Z_{itz}^{age=-1}$ & 1.055***& &    1.052***& &    1.070***& &    1.055***&  \\
          &  (0.099)   & &  (0.098)   & &  (0.104)   & &  (0.098)   &  \\

$UE_{itz}^{age=-1}$ & &   -0.021   & &   -0.004   & &   -0.024   & &    0.112   \\
          & &  (0.047)   & &  (0.006)   & &  (0.028)   & &  (0.120)   \\

F statistic &   114.52   & &   114.40   & &   105.22   & &   114.88   &  \\ [6pt]
\multicolumn{5}{l}{\emph{Panel B: Local unemployment at age 0}}\\
$Z_{itz}^{age=0}$&    1.072***& &    1.068***& &    1.076***& &    1.072***&  \\
          &  (0.086)   & &  (0.085)   & &  (0.094)   & &  (0.086)   &   \\

$UE_{itz}^{age=0}$ & &   -0.015   & &    0.003   & &    0.049** & &   -0.137   \\
          & &  (0.044)   & &  (0.006)   & &  (0.024)   & &  (0.106)   \\
F statistic &   155.90   & &   157.56   & &   130.18   &  &   157.22   & \\  [6pt]
\multicolumn{5}{l}{\emph{Panel C: Local unemployment at age 1}}\\
$Z_{itz}^{age=1}$  &    1.046***& &    1.048***& &    1.068***& &    1.048***&    \\
          &  (0.080)   &  &  (0.079)   & &  (0.094)   & &  (0.080)   &    \\

$UE_{itz}^{age=1}$& &   -0.008   & &   -0.002   & &    0.005   & &   -0.073   \\
          & &  (0.039)   & &  (0.006)   & &  (0.022)   & &  (0.114)   \\

F statistic&   170.93   & &   177.21   & &   129.57   & &   170.18   &   \\

\hline
Observations       & 109,842  & 109,842& 97,406     & 97,406  & 54,597    & 54,597   & 110,725 & 110,725   \\

\hline\hline

\end{tabularx}
\begin{tablenotes}[para,flushleft]
     \footnotesize{Notes: Education is measured in years. Wages are measured as log hourly wages imputed using individuals' occupations. Fluid intelligence is standardised within the estimation sample to have mean 0, standard deviation 1. Height is individuals' standing height (in cm). We include controls for gender, month-of-birth and year-of-birth fixed effects, area-of-birth fixed effects, and area-specific trends. Standard errors in parentheses are clustered at area level. \sym{*} \(p<0.1\), \sym{**} \(p<0.05\), \sym{***} \(p<0.01\)}
\end{tablenotes}
\end{threeparttable}
\end{adjustbox}
\end{table}

%% file: tables/1000_days_first.tex
\begin{table}[htbp]\centering
\def\sym#1{\ifmmode^{#1}\else\(^{#1}\)\fi}
\caption{Effect of instrument on area unemployment during the first 1,000 days --- First stage for joint estimation\label{tab: 1000 days first}}
\begin{adjustbox}{max width=\textwidth}
\begin{threeparttable}
\begin{tabular}{l*{6}{c}}
\hline\hline
                &\multicolumn{1}{c}{(1)}&\multicolumn{1}{c}{(2)}&\multicolumn{1}{c}{(3)}&\multicolumn{1}{c}{(4)}&\multicolumn{1}{c}{(5)}&\multicolumn{1}{c}{(6)}\\
		&\multicolumn{1}{c}{$UE_{itz}^{age=-1}$}&\multicolumn{1}{c}{$UE_{itz}^{age=0}$}&\multicolumn{1}{c}{$UE_{itz}^{age=1}$}&\multicolumn{1}{c}{$UE_{itz}^{age=-1}$}&\multicolumn{1}{c}{$UE_{itz}^{age=0}$}&\multicolumn{1}{c}{$UE_{itz}^{age=1}$}\\
\hline
\textbf{Panel A}    &\multicolumn{3}{c}{\textbf{Education (years)}}&\multicolumn{3}{c}{\textbf{Log hourly wage}} \\

\hline
$Z_{itz}^{age=-1}$ &    1.492***&    0.125   &   -0.124***&    1.488***&    0.125   &   -0.123***\\
                &  (0.135)   &  (0.082)   &  (0.043)   &  (0.136)   &  (0.081)   &  (0.043)   \\
$Z_{itz}^{age=0}$ &    0.080   &    1.480***&    0.105   &    0.075   &    1.474***&    0.105   \\
                &  (0.050)   &  (0.105)   &  (0.074)   &  (0.049)   &  (0.104)   &  (0.072)   \\
$Z_{itz}^{age=1}$ &    0.011   &    0.145***&    1.441***&    0.007   &    0.146***&    1.443***\\
                &  (0.081)   &  (0.041)   &  (0.101)   &  (0.078)   &  (0.040)   &  (0.099)   \\
F statistic & 189.94     & 240.60     & 206.22     & 182.34     & 232.60     & 211.64 \\	
\hline 
Observations    &   \multicolumn{3}{c}{109,842}&    \multicolumn{3}{c}{97,406}    \\

\hline
\textbf{Panel B}     &\multicolumn{3}{c}{\textbf{Fluid intelligence}}&\multicolumn{3}{c}{\textbf{Height (cm)}} \\
    
\hline

$Z_{itz}^{age=-1}$&    1.526***&    0.099   &   -0.138***&    1.492***&    0.125   &   -0.124***\\
                &  (0.125)   &  (0.102)   &  (0.048)   &  (0.135)   &  (0.082)   &  (0.043)   \\
$Z_{itz}^{age=0}$&    0.072   &    1.537***&    0.084   &    0.079   &    1.480***&    0.106   \\
                &  (0.046)   &  (0.111)   &  (0.104)   &  (0.050)   &  (0.105)   &  (0.073)   \\
$Z_{itz}^{age=1}$&   -0.074   &    0.092** &    1.512***&    0.010   &    0.145***&    1.443***\\
                &  (0.105)   &  (0.037)   &  (0.113)   &  (0.080)   &  (0.041)   &  (0.102)   \\
F statistic & 188.39     & 209.16     & 219.91     & 188.56     & 236.61     & 202.25 \\	
\hline 
Observations    &   \multicolumn{3}{c}{54,597}&    \multicolumn{3}{c}{110,725}    \\
\hline\hline

\end{tabular}
\begin{tablenotes}[para,flushleft]
     \footnotesize{Notes: The instrument $Z_{itz}^{age=a}$ is the product of pre-determined area-level industry shares at age a and national industry-specific unemployment rates at age a. We include controls for gender, month-of-birth and year-of-birth fixed effects, area-of-birth fixed effects, and area-specific trends. Variation in the estimates between the four outcomes of interest are driven by changes in the sample size only. Standard errors in parentheses are clustered at area level.  \sym{*} \(p<0.1\), \sym{**} \(p<0.05\), \sym{***} \(p<0.01\)}
\end{tablenotes}
\end{threeparttable}
\end{adjustbox}
\end{table}

%% file: tables/1000_days_bartik.tex
\begin{table}[htbp]\centering
\def\sym#1{\ifmmode^{#1}\else\(^{#1}\)\fi}
\caption{IV estimates of the effect of early life economic conditions on individuals' human capital and health outcomes during the first 1,000 days --- Joint estimation \label{tab: 1000 days iv together}}
\begin{threeparttable}
\begin{tabular}{l*{4}{c}}
\hline\hline
                           &\multicolumn{1}{c}{(1)}&\multicolumn{1}{c}{(2)}&\multicolumn{1}{c}{(3)}&\multicolumn{1}{c}{(4)}\\
                &\multicolumn{1}{c}{\makecell{Education \\ (years)}}&\multicolumn{1}{c}{\makecell{Log \\ hourly wages}}&\multicolumn{1}{c}{\makecell{Fluid \\ intelligence}}&\multicolumn{1}{c}{\makecell{Height \\ (cm)}}\\
\hline
$UE_{itz}^{age=-1}$&     -0.023   &   -0.008   &   -0.043   &    0.173   \\
                &  (0.048)   &  (0.006)   &  (0.036)   &  (0.140)   \\
$UE_{itz}^{age=0}$&   -0.00003   &    0.004   &    0.056** &   -0.148   \\
                &  (0.047)   &  (0.006)   &  (0.027)   &  (0.115)   \\
$UE_{itz}^{age=1}$&   -0.017   &   -0.006   &   -0.016   &   -0.021   \\
                &  (0.041)   &  (0.006)   &  (0.028)   &  (0.133)   \\

F statistic          &    50.31   &    50.40   &    51.79   &    50.27   \\
\hline
Observations    &   109,842   &    97,406   &    54,597   &   110,725   \\

\hline\hline

\end{tabular}
\begin{tablenotes}[para,flushleft]
     \footnotesize{Notes: Education is measured in years. Wages are measured as log hourly wages imputed using individuals' occupation. Fluid intelligence is standardised within the estimation sample to have mean 0, standard deviation 1. Height is individuals' standing height (in cm). We include controls for gender, month-of-birth and year-of-birth fixed effects, area-of-birth fixed effects, and area-specific trends. Standard errors in parentheses are clustered at area level. \sym{*} \(p<0.1\), \sym{**} \(p<0.05\), \sym{***} \(p<0.01\)}
\end{tablenotes}
\end{threeparttable}
\end{table}

%% file: tables/Reduced_form_estimated.tex
\begin{table}[htbp]\centering
\def\sym#1{\ifmmode^{#1}\else\(^{#1}\)\fi}
\caption{Effect of area level instrument on individuals' human capital and health outcomes -- Reduced form in the estimation sample \label{tab: reduced form estimation sample}}
\begin{threeparttable}
\begin{tabular}{l*{4}{c}}
\hline\hline
                &\multicolumn{1}{c}{(1)}&\multicolumn{1}{c}{(2)}&\multicolumn{1}{c}{(3)}&\multicolumn{1}{c}{(4)}\\
                &\multicolumn{1}{c}{\makecell{Education \\ (years)}}&\multicolumn{1}{c}{\makecell{Log \\ hourly wages}}&\multicolumn{1}{c}{\makecell{Fluid \\ intelligence}}&\multicolumn{1}{c}{\makecell{Height \\ (cm)}}\\
\hline
\multicolumn{5}{l}{\emph{Panel A: Weighted industry unemployment during pregnancy}} \\

$Z_{itz}^{age=-1}$ &   -0.024   &   -0.006   &   -0.034   &    0.218   \\
          &  (0.064)   &  (0.009)   &  (0.040)   &  (0.162)   \\

\multicolumn{5}{l}{\emph{Panel B: Weighted industry unemployment at age 0}}\\

$Z_{itz}^{age=0}$ &   -0.011   &    0.002   &    0.068*  &   -0.160   \\
          &  (0.065)   &  (0.009)   &  (0.035)   &  (0.154)   \\

\multicolumn{5}{l}{\emph{Panel C: Weighted industry unemployment at age 1}}\\

$Z_{itz}^{age=1}$&   -0.013   &   -0.005   &    0.010   &   -0.150   \\
          &  (0.057)   &  (0.008)   &  (0.032)   &  (0.162)   \\
\hline
Observations&  109,842   &   97,406   &   54,597   &  110,725   \\

\hline\hline

\end{tabular}
\begin{tablenotes}[para,flushleft]
     \footnotesize{Notes: Education is measured in years. Wages are measured as log hourly wages imputed using individuals' occupation. Fluid intelligence is standardised within the estimation sample to have mean 0, standard deviation 1. Height is individuals' standing height (in cm). The instrument $Z_{itz}^{age=a}$ is the product of pre-determined area-level industry shares at age a and national industry-specific unemployment rates at age a. We include controls for gender, month-of-birth and year-of-birth fixed effects, area-of-birth fixed effects, and area-specific trends. Standard errors in parentheses are clustered at area level. \sym{*} \(p<0.1\), \sym{**} \(p<0.05\), \sym{***} \(p<0.01\)}
\end{tablenotes}
\end{threeparttable}
\end{table}

%% file: tables/Reduced_form_whole_sample.tex
\begin{table}[htbp]\centering
\def\sym#1{\ifmmode^{#1}\else\(^{#1}\)\fi}
\caption{Effect of district-level instrument on individuals' human capital and health outcomes -- Reduced form in the sample of all districts in England and Wales \label{tab: reduced form whole sample}}
\begin{threeparttable}
\begin{tabular}{l*{4}{c}}
\hline\hline
                &\multicolumn{1}{c}{(1)}&\multicolumn{1}{c}{(2)}&\multicolumn{1}{c}{(3)}&\multicolumn{1}{c}{(4)}\\
                &\multicolumn{1}{c}{\makecell{Education \\ (years)}}&\multicolumn{1}{c}{\makecell{Log \\ hourly wages}}&\multicolumn{1}{c}{\makecell{Fluid \\ intelligence}}&\multicolumn{1}{c}{\makecell{Height \\ (cm)}}\\
\hline
\multicolumn{5}{l}{\emph{Panel A: Weighted industry unemployment during pregnancy}} \\

$Z_{itd}^{age=-1}$ &   -0.007   &   -0.002   &    0.013   &    0.115   \\
          &  (0.057)   &  (0.009)   &  (0.037)   &  (0.147)   \\

\multicolumn{5}{l}{\emph{Panel B: Weighted industry unemployment at age 0}}\\

$Z_{itd}^{age=0}$ &   -0.012   &    0.002   &    0.067*  &   -0.110   \\
          &  (0.058)   &  (0.009)   &  (0.038)   &  (0.154)   \\

\multicolumn{5}{l}{\emph{Panel C: Weighted industry unemployment at age 1}}\\

$Z_{itd}^{age=1}$ &   -0.045   &   -0.008   &   -0.023   &   -0.185   \\
          &  (0.058)   &  (0.008)   &  (0.035)   &  (0.158)   \\

\hline
Observations&  151,560   &  134,799   &   73,760   &  152,624   \\

\hline\hline

\end{tabular}
\begin{tablenotes}[para,flushleft]
     \footnotesize{Notes: Education is measured in years. Wages are measured as log hourly wages imputed using individuals' occupations. Fluid intelligence is standardised within the estimation sample to have mean 0, standard deviation 1. Height is individuals' standing height (in cm). The instrument $Z_{itd}^{age=a}$ is the product of pre-determined district-level industry shares at age a and national industry-specific unemployment rates at age a. We include controls for gender, month-of-birth and year-of-birth fixed effects, district-of-birth fixed effects, and district-specific trends. Standard errors in parentheses are clustered at district level. \sym{*} \(p<0.1\), \sym{**} \(p<0.05\), \sym{***} \(p<0.01\)}

\end{tablenotes}
\end{threeparttable}
\end{table}

%% file: tables/gwas_descriptives.tex
\begin{table}
\caption{\label{tab:gwas-descriptives}GWAS samples -- Descriptives}
\centering
\begin{threeparttable}
\begin{tabular}[t]{lrrr}
\hline\hline
\multicolumn{1}{c}{} & \multicolumn{1}{c}{(1)} & \multicolumn{1}{c}{(2)} & \multicolumn{1}{c}{(3)} \\
\multicolumn{1}{c}{} & \multicolumn{1}{c}{Obs.} & \multicolumn{1}{c}{Mean} & \multicolumn{1}{c}{Std. dev.}\\
\midrule
Education (years)    & 338,248      & 13.018   &  2.371   \\
Log hourly wages     & 184,585      & 2.625    &  0.354   \\
Fluid intelligence   & 143,117      & 6.137    &  2.097   \\
Height (cm)          & 341,080      & 168.311  &  9.217   \\
\hline\hline
\end{tabular}
\begin{tablenotes}[para,flushleft]
\footnotesize{Notes: Columns correspond to (1) Number of observations in the GWAS sample. (2) Sample mean in the GWAS sample. (3) Sample standard deviation in the GWAS sample.}
\end{tablenotes}
\end{threeparttable}
\end{table}

%% file: tables/pgs_predictive_power.tex
\begin{table}\centering
\def\sym#1{\ifmmode^{#1}\else\(^{#1}\)\fi}
\caption{\label{tab:pgs_predictive}Predictive power of the polygenic indices}
\centering
\setlength{\tabcolsep}{8pt}
\begin{threeparttable}
\begin{tabular}[t]{lrrrr}
\hline\hline
\multicolumn{1}{c}{\em{}} & \multicolumn{4}{c}{\em{Dependent variable:}} \\
\cmidrule(l{3pt}r{3pt}){2-5}
\multicolumn{1}{c}{} & \multicolumn{1}{c}{(1)} & \multicolumn{1}{c}{(2)} & \multicolumn{1}{c}{(3)} & \multicolumn{1}{c}{(4)} \\
\multicolumn{1}{c}{ } & \multicolumn{1}{c}{\begin{tabular}[c]{@{}c@{}} Education \\ (years)\end{tabular}}& \multicolumn{1}{c}{\begin{tabular}[c]{@{}c@{}}Log hourly \\ wages\end{tabular}} & \multicolumn{1}{c}{\begin{tabular}[c]{@{}c@{}}Fluid \\ intelligence \end{tabular}} & \multicolumn{1}{c}{\begin{tabular}[c]{@{}c@{}} Height \\ (cm) \end{tabular}} \\
\midrule
\addlinespace[0.3em]
PGI & $0.747^{ *** }$ & $0.060^{ *** }$ & $0.301^{ *** }$ & $3.882^{ *** }$ \\
 & $(0.014)$ & $(0.001)$ & $(0.004)$ & $(0.019)$ \\
\addlinespace[0.75em]
Observations &     104,051         &      92,058         &      51,073         &     104,404 \\
Mean dep. var. & 13.3 & 2.6 & 0 & 169.6 \\
$R^2$ & 0.112 & 0.149 & 0.090 & 0.680 \\
Incremental $R^2$ & 0.104 & 0.027 & 0.082 & 0.168 \\
\hline\hline
\end{tabular}
\begin{tablenotes}[para,flushleft]
    \footnotesize{Notes: Columns are for (1) educational attainment in years, (2) logarithm of hourly wage, (3) standardised fluid intelligence score, (4) adult height in centimetres. Regresses each outcome onto its corresponding polygenic index as well as the first 10 genetic principal components and an indicator for gender. Polygenic indices are standardised to zero mean and unit variance. Incremental R$^2$ is the increase in R$^2$ when the polygenic index is included as a covariate.  \sym{*} \(p<0.1\), \sym{**} \(p<0.05\), \sym{***} \(p<0.01\)}
\end{tablenotes}
\end{threeparttable}
\end{table}